\documentclass[twocolumn,aps,prl,amsmath,amssymb,groupofadress,showpacs]{revtex4-1}

\usepackage{graphicx}
\usepackage{graphics}
\usepackage{dcolumn}
\usepackage{bm}
\usepackage{color}
\usepackage{soul}
\usepackage{textcomp}
\usepackage{soul}

\usepackage{multirow}
\usepackage{setspace}
\usepackage[mathlines]{lineno}
\usepackage{textcomp}
\usepackage{mathcomp}

\usepackage{microtype}
\usepackage[breaklinks=true,hyperindex=true,
pdftitle={Entangled Photonic-Nuclear Molecular Dynamics of LiF in Quantum Optical Cavities},
colorlinks=true,pagebackref=false,citecolor=blue,plainpages=false,pdfpagelabels,
linkcolor=blue,urlcolor=blue]{hyperref}

\begin{document}

\title{
Entangled Photonic-Nuclear Molecular Dynamics of LiF in Quantum Optical Cavities
}

\author{Johan F. Triana}
\affiliation{Grupo de F\'{\i}sica At\'omica y Molecular, Instituto de F\'{\i}sica, Universidad de Antioquia, Medell\'{\i}n, Colombia}
\author{Daniel Pel\'aez}
\affiliation{Laboratoire de Physique des Lasers, Atomes et Mol\'ecules (PhLAM), Unit\'e Mixte de Recherche (UMR) 8523, Universit\'e Lille 1, B\^at. P5,Villeneuve d'Ascq Cedex, France}
\author{Jos\'e Luis Sanz-Vicario}
\affiliation{Grupo de F\'{\i}sica At\'omica y Molecular, Instituto de F\'{\i}sica, Universidad de Antioquia, Medell\'{\i}n, Colombia}
\email{jose.sanz@udea.edu.co}
%

\begin{abstract}
 
The quantum photodynamics of a simple diatomic molecule with a permanent dipole immersed within an optical cavity containing a quantized radiation field
is studied in detail. The chosen molecule under study, lithium fluoride (LiF), is characterized by the presence of an avoided crossing 
between the two lowest $^1\Sigma$ potential energy curves (covalent-ionic diabatic crossing). Without field, after prompt excitation from the ground state $1\; ^1\Sigma$, 
the excited nuclear wave packet moves back and forth in the upper $2\; ^1\Sigma$ state, but in the proximity of the avoided crossing, the non-adiabatic coupling
transfers part of the nuclear wave packet to the lower $1\; ^1\Sigma$ state, which eventually leads to dissociation.
The quantized field of a cavity also induces an additional light crossing in the modified dressed potential energy curves with similar transfer properties.  
To understand the entangled photonic-nuclear dynamics we solve the time dependent Schr\"odinger equation by using the multiconfigurational time dependent 
Hartree method (MCTDH). The single mode quantized field of the cavity is represented in the coordinate space instead of in the Fock space, which 
allows us to deal with the field as an additional vibrational mode within the MCTDH procedure on equal footing. We prepare the cavity with different quantum
states of light, namely, Fock states, coherent states and squeezed coherent states. Our results reveal pure quantum light effects on the molecular photodynamics 
and the dissociation yields of LiF,  which are quite different from the light-undressed case and that cannot be described in general by a semiclassical approach
using classical electromagnetic fields.
\end{abstract}

\date{\today}

\maketitle

\section{Introduction}

The interaction of molecules with light is a cornerstone in the development of molecular quantum mechanics. Along the recent decades the two different
areas of i) quantum optics and ii) semiclassical molecular dynamics have evolved almost independently. The former is mostly interested in new emerging 
phenomena rising from the quantum nature of light, using model Hamiltonians for open quantum systems with a minimal number of states (qubits) \cite{Gerry2005,Walls2009}
The latter focuses  on the molecular structure and dynamics under conditions in which the quantum description of light is not required and the interaction with cw radiaton or 
pulsed lasers employs semiclassical approximations, which incorporate the classical fields from Maxwell electrodynamics \cite{Tannor2007}. In this case, methods of
solution involving a large number of vibronic states is the rule.  A no-man's land located in between is still quite unexplored: 
ab initio polaritonic molecular photodynamics, i.e., many-state molecular dynamics (with a full vibronic description)  along with a full 
quantum description of light. Of course, the physical situation with the conditions of application of such a full quantum theory must be met. 
A potential case is represented by molecules passing by or confined within optical cavities \cite{Walther2006,Haroche2006}.

Since the seminal works by Zewail \cite{Zewail1994,Tannor2007}, it was clear that molecular photoreactions can be controlled with the use of ultrashort laser fields. 
The area of photodynamics with ultrashort laser pulses, down to the regimen of attosecond, has developed rapidly in the last years \cite{Palacios2015}. 
However, in these recent theoretical studies the molecule is subject to classical fields. When an atom or molecule is confined within a quantum cavity, the light must 
be considered as a set of quantized modes and the tools of quantum optics apply. There is a vast literature  on cavity quantum electrodynamics 
(cavity QED) specially dedicated to atom-cavity interactions and control \cite{Walther2006,Haroche2006}. The study of molecular photodynamics in quantum cavities 
is more scarce but nevertheless quite intense at present  \cite{Galego2015,Herrera2016,Kowalewski2016,Kowalewski2017,Flick2017,Kowalewski2017a,Csehi2017a,Csehi2017b,Galego2017}.
  
Molecules in cavities may reach a strong coupling regime in the light-matter interaction, resulting in structure and dynamics better understood 
under consideration of light-modified (or dressed) potential energy surfaces. Strong coupling regime is favored by the small volume of the cavity, 
hence the potential use of confinements in micro or nano-cavities. The typical treatment in quantum optics for  these coupled systems is 
based in models like the two-state Jaynes-Cummings model \cite{Gerry2005}. The eigenstates of the Jaynes-Cummings model (which assumes the 
RWA approximation) are called dressed (by the light) states. From the eigenvalues of the dressed states one can extract the new dressed potential 
energy surfaces of the molecule, that in the molecular case display new avoided crossings or conical intersections induced by the light. This phenomena 
of light induced crossings (LIC) and conical intersections (LICI) is well described in the literature \cite{Gabor2012}, Also, most of the interesting molecular 
photo-reactivity is produced through electronic excited states. The molecular wave packets promoted to these excited states find different light-modified 
landscapes for its subsequent dynamics,  thus opening new channels for photophysical or photochemical reactions. A goal in control theory is to select 
specific reaction paths by manipulating the evolution of the wave packet in the manifold of excited states.  Note that in this strong coupling regime the 
molecular modes of motion and the light cavity modes become entangled in a combined non-separable light-matter wave function. 

In this work we address the inner workings of the photodynamics of a diatomic molecule like LiF inside a quantized optical cavity.
Previous recent papers \cite{Kowalewski2016,Kowalewski2017,Csehi2017a,Csehi2017b} have studied the heavier molecule NaI. From the theoretical point of view, 
since the LiF is lighter than NaI its  nuclear dynamics is faster. Then similar physics occurring in NaI also takes place in LiF but in a shorter time, which is a 
computational advantage.  We show in this work that the molecular cavity photodynamics of LiF may drastically change upon the quantum 
field present in the cavity. In addition, contrary to previous statements \cite{Csehi2017a,Csehi2017b} we believe that our results indicate that the fully quantum 
nature of the radiation is expressed in features that cannot be reproduced whatsoever using a classical light.

\section{Theory}

We assume an optical cavity with a single mode confined radiation with frequency $\omega_c$. 
In basic textbooks in quantum optics \cite{Gerry2005,Walls2009} it is shown that the quantized field can be 
represented by a quantum harmonic oscillator (HO) with mass unity  for each radiation mode given by its frequency $\omega_c$. The field Hamiltonian is simply 
$\hat{H}_{\rm Field} = \hbar \omega_c(\hat{a}^\dag \hat{a} + 1/2)$. The electric field operator
for a single mode (propagated along the $q$ direction but polarized in the $z$ direction with unit vector $\hat{\varepsilon}_z$) reads
\begin{equation}
\hat{E}_z (q) =  E_0 (\hat{a} + \hat{a}^\dag) \sin (kq) \hat{\varepsilon}_z ,
\end{equation}
where $E_0 = \sqrt{ \frac{\hbar\omega_c}{V \epsilon_0}}$ is the field amplitude and $V$ is the volume of the cavity.  The quantum version of the two-state $\{ | g\rangle, |e\rangle \}$ Rabi model with dipolar coupling $\hat{\mu}_{eg}$ is named Jaynes-Cummings model \cite{Gerry2005,Walls2009}, whose Hamiltonian before the RWA approximation reads
\begin{align}
\label{eq:Hamiltot}
\hat{H} & = \hat{H}_{\text{Mol}} + \hat{H}_{\rm Field} + \hat{H}^{I}_{\rm MF}  \nonumber \\
             & = \frac{1}{2} \hbar \omega_0 \hat{\sigma}_3 +   \hbar \omega_c(\hat{a}^\dag \hat{a} + 1/2) + 
\hbar \lambda  (\hat{a} + \hat{a}^\dag) (\hat{\sigma}_+  +  \hat{\sigma}_-)
\end{align}
where $\omega_0 = (E_e -E_g)/\hbar$ is the natural frequency, the coupling factor $\lambda$ is defined as $\lambda =  \hat{\mu}_{ge; z} E_0 \sin (kq)/\hbar$, $\hat{\sigma}_+ = |e\rangle \langle g |$, $\hat{\sigma}_- =\hat{\sigma}^\dag_+ $ 
are transition operators, and $\hat{\sigma}_3 = |e\rangle \langle  e| -|g \rangle \langle g |$ is the inversion operator. The Jaynes-Cumings model admits a general 
time dependent solution in the interaction picture (and within the RWA) in the entangled form $|\Psi(t)\rangle = \sum^M_{n=0} [a_n(t) |g,n\rangle + b_n(t) |e,n\rangle  ]$ \cite{Gerry2005},
for any two-state bare system prepared initially as a superposition $|\psi (0)\rangle_{\rm Mol} = 
C_g | g\rangle + C_e |e \rangle$ and any field represented in the Fock basis of photons, i.e., $|\psi(0) \rangle_{\rm Field} = \sum_{n=0}^M C_n |n\rangle$ (can be a Fock state, a coherent
state, a vacuum squezeed state, a squeezed coherent state, etc).  In the present molecular problem, the number $M$, the number of states in the superposition in the Fock space, 
can be large, which makes this route of solution very demanding. To circumvent the difficulties
related to the representation of light in a Fock basis one can relay in the direct representation of the the cavity light mode as a quantum harmonic 
oscillator of mass unity in coordinate space \cite{Kowalewski2017}. Since $ (\hat{a} + \hat{a}^\dag) = \sqrt{\frac{2\omega_c}{\hbar}} \hat{x}$ the interaction term in the Jaynes-Cummings Hamiltonian is expressed
as $\hat{\mu}_{ge} \chi \omega_c \sqrt{2\hbar}   \hat{x}  (\hat{\sigma}_+  +  \hat{\sigma}_-) $, with $\chi= \sqrt{\frac{1}{\hbar V  \epsilon_0}} \sin(kq)$.

All these ideas can be applied to the solution of the two-state molecular photodynamics of a linear molecular like LiF.
In this work we assume for simplicity that LiF molecules within the cavity are oriented along the polarization axis of the cavity radiation field, so that only $\Sigma-\Sigma$ radiative transitions are allowed. In the absence of radiation the usual method of solution
is to expand the total vibronic wave function for a diatomic molecule in terms of the complete set of adiabatic electronic states. In this two-state model, they are named $\{ |g\rangle , |e\rangle  \}
$ (after ground and excited),  whose energies $E_g(R)$ and $E_e(R)$, respectively, correspond to the potential energy curves (PEC) as functions of the internuclear distance $R$. Thus the 
total wave packet has the  form $|\Psi(t)\rangle = \varphi_g (R,t) |g\rangle + \varphi_e(R,t) |e\rangle$, where $\varphi_i(R,t)$ $(i=e,g)$ represents 
the time dependent nuclear wave packet moving in its respective PEC. When the interaction with the quantized cavity is also included the ansatz takes instead the form
\begin{equation}
\label{eq:tdansatz}
|\Psi(t)\rangle = \varphi_g (R,x,t) |g\rangle + \varphi_e(R,x,t) |e\rangle,
\end{equation}
where now $\varphi_i (R,x,t)$ represent entangled wave packets for the joint vibrational and cavity photonic mode dynamics. 
In order to solve the photodynamics of the molecule in the cavity the time dependent Schr\"odinger equation 
(TDSE),  $[ i \hbar \frac{\partial}{\partial t} - \hat{H}] | \Psi(t) \rangle=0$, must be solved.
In this particular problem the total Hamiltonian $\eqref{eq:Hamiltot}$ must include the molecular Hamiltonian for the diatom LiF, i.e., $\hat{H}_{\rm Mol} =  \hat{T}_{\rm N} + \hat{H}_{\rm el}$.
Also, the adiabatic basis of electronic states are eigenfunctions of the electronic Hamiltonian, i.e., $[\hat{H}_{\rm el} -E_g(R)] |g\rangle =0$ and   $[\hat{H}_{\rm el} -E_e(R)] |e\rangle =0$.
With  these premises, if we insert the ansatz  \eqref{eq:tdansatz} into the TDSE we obtain an equation in matrix form
\begin{widetext}
\begin{eqnarray}\label{eq:coupledTDSE}
i \hbar \frac{d}{dt} &\left[ 
\begin{array}{c}
 \varphi_g (R,x,t)  \\
 \varphi_e (R,x,t)  
\end{array}
\right]  = \left\{ \left(
\begin{array}{cc}
    -\frac{\hbar^2}{2\mu}  \frac{\partial^2}{\partial R^2}  + E_g(R) -\frac{\hbar^2}{2}\frac{\partial^2}{\partial x^2} + \frac{1}{2}\omega^2_c x^2  & 0  \\
0 &   -\frac{\hbar^2}{2\mu}  \frac{\partial^2}{\partial R^2} + E_e(R)   -\frac{\hbar^2}{2}\frac{\partial^2}{\partial x^2}  + \frac{1}{2}\omega^2_c x^2  \end{array}
\right) \right. \nonumber \\
& \left. + \left(
\begin{array}{cc}
      \chi \omega_c \sqrt{2\hbar}  \mu_{gg}(R)   x       &   \chi \omega_c \sqrt{2\hbar}  \mu_{ge}(R)   x \\
       \chi \omega_c \sqrt{2\hbar}  \mu_{eg}(R)  x  &    \chi \omega_c \sqrt{2\hbar}  \mu_{ee}(R)   x       
\end{array}
\right) + {\mathbf C} (R) \right\}
\left[ \begin{array}{c}
 \varphi_g (R,x,t)   \\
 \varphi_e (R,x,t)  
 \end{array}
\right],
\end{eqnarray}
%
where $\mu$ is the reduced nuclear mass, $\mu_{ij}$ are the molecular electronic dipole moments and the matrix elements for $\mathbf{C}$,
%
\begin{equation}
C_{ij}(R)=-\frac{\hbar^{2}}{2\mu}\left[ 2\left\langle i\left|\frac{\partial}{\partial R}\right| j \right\rangle
(R) \frac{\partial}{\partial R} + \left\langle i\left|\frac{\partial^{2}}{\partial R^{2}}\right| j \right\rangle
(R) \right] = -\frac{\hbar^{2}}{2\mu}\left[ 2f_{ij}(R)\frac{\partial}{\partial R} + h_{ij}(R)\right]
\end{equation}
\end{widetext}
correspond to the well known non adiabatic couplings. In compact matrix form this Hermitian term reads 
$\boldsymbol{\mathrm{C}}=-\frac{\hbar}{2\mu}\left[2\boldsymbol{\mathrm{F}}\frac{\mathrm{d}}{\mathrm{d}R}+\boldsymbol{\mathrm{H}} \right]$.
The term ${\bf H}$ with second derivatives with respect to $R$ is more difficult to compute in general, but with an expansion in two electronic states is simple. If one inserts
the completeness $\hat{1} = \sum_{i={g,e}} | i\rangle \langle i |$ in $h_{ij}$ one arrives to the identity 
$\boldsymbol{\mathrm{H}}=\boldsymbol{\mathrm{F}}^{2}+\frac{\mathrm{d}\boldsymbol{\mathrm{F}}}{\mathrm{d}R}$. In 
our particular case

\begin{equation} \label{eq:Fcoup}
\mathbf{F}=
\begin{pmatrix}
0                & f_{ge} \\
-f_{ge} & 0
\end{pmatrix}
\end{equation}
is an anti-Hermitian matrix, so that
\begin{equation}\label{eq:Hcoup}
\mathbf{H}=
\begin{pmatrix}
-|f_{ge}|^2              & \frac{{\rm d} f_{ge}}{{\rm d}R}  \\
-\frac{{\rm d} f_{ge}}{{\rm d} R} & -|f_{ge}|^2 
\end{pmatrix}.
\end{equation}
Here $\mathbf{H}$ is also an anti-hermitian matrix but, compensated with the term $2 \mathbf{F} \frac{{\rm d}}{{\rm d}R}$, the full matrix $\mathbf{C}$ is Hermitian and the propagation preserves unitarity.
In short, only one non-adiabatic coupling, $f_{ge}(R)$, must be computed.

The system of dynamic coupled equations \eqref{eq:coupledTDSE} can be solved in different ways. We choose to solve it by using the multiconfigurational time-dependent Hartree method
(MCTDH) \cite{Meyer1990,Beck2000}, for which a computational toolkit is made available for free upon request \cite{MeyerMCTDH2007}. In brief, the photonic-nuclear wave packets $\varphi_{g,e} (R,x,t)$ (with two degrees of freedom) are expanded in terms of time-dependent single particle functions $\phi^{\kappa}_{j_\kappa} (Q_{\kappa},t)$ for each coordinate $Q_{\kappa}$. In our case
the MCTDH ansatz takes the form 
\begin{equation}
\varphi_{g,e}(R,x,t) = \sum^{n_R}_{j_{R}=1}  \sum^{n_x}_{j_x=1} A_{j_R. j_x} (t) \phi^{R}_{j_R} (R) \phi^{x}_{j_x} (x) .
\end{equation}
The single particle functions are represented in terms of a sin-DVR primitive basis set for the molecular coordinate $R$ and
HO-DVR primitive basis set for the cavity mode coordinate $x$ as follows
\begin{equation}
\phi^{q}_{j_q} (q,t) = \sum^{N_q}_{i=1} C^{(q)}_{j_q i} (t) \chi^{(q)}_i (q); \quad q=R,x.
\end{equation}
In the MCTDH method of solution it is necessary to obtain the time dependent expansion coefficients $A_{j_R. j_x} (t)$ as well
as the time-dependent single particle functions to build the total wave packet. The MCTDH equations of motion for the set
coefficients $A_J(t)$ and single particle funcions  are \cite{Beck2000,MCTDHBook2009}
\begin{align}
& i \dot{A}_{\bf J} = \sum_{\bf L} \langle \Phi_{\bf J} | H | \Phi_{\bf L} \rangle A_{\bf L} \nonumber \\
& i \dot{{ \bm \phi}}^{(\kappa)}= (1-P^{(\kappa)}) ({\bm \rho}^{(\kappa)} )^{-1} \langle  {\bf H} \rangle^{(\kappa)} {\bm \phi}^{(\kappa)}.
\end{align}
These equations are implemented in the MCTDH software package \cite{MeyerMCTDH2007}. 
MCTDH package requires the input of the potential energy surfaces (PES) involved. In this case, the $(R,x)$-PES for the
ground state $g$ is  $E_g(R) + \frac{1}{2}\omega^2_c x^2  + \chi \omega_c \sqrt{2\hbar} \mu_{gg}(R) x$
and for the excited state is  
$E_e(R) + \frac{1}{2}\omega^2_c x^2  + \chi \omega_c \sqrt{2\hbar} \mu_{ee}(R) x$.
It means that these PES contain the molecular PECs of LiF ($1 ^1\Sigma$ for $g$ and $2 ^1\Sigma$ for $e$) plus the radiation 
HO potential and the diagonal terms of the interaction with the dipole moments $\mu_{gg}$ and $\mu_{ee}$ that 
cannot be discarded in this polar molecule. The MCTDH provides, among many other observables, 
 the time-dependent evolution of populations for the $g$ and $e$ light-matter entangled states, the dissociation probabilities computed 
 with the flux moving across a complex absorbing potential and, more interestingly, the separate entangled wave packets $\varphi_g (R,x,t)$ and $\varphi_e (R,x,t)$,
 evolving in the $g$ and $e$ PES, that we call $g$-wave packet  and $e$-wave packet in this work. The temporal analysis of these
  wave packets help us to reveal the underlying mechanism in this complex entangled photonic-nuclear dynamics.

\section{Results and discussion}

\begin{figure}[h!]
\centering
\includegraphics[width=0.5\textwidth]{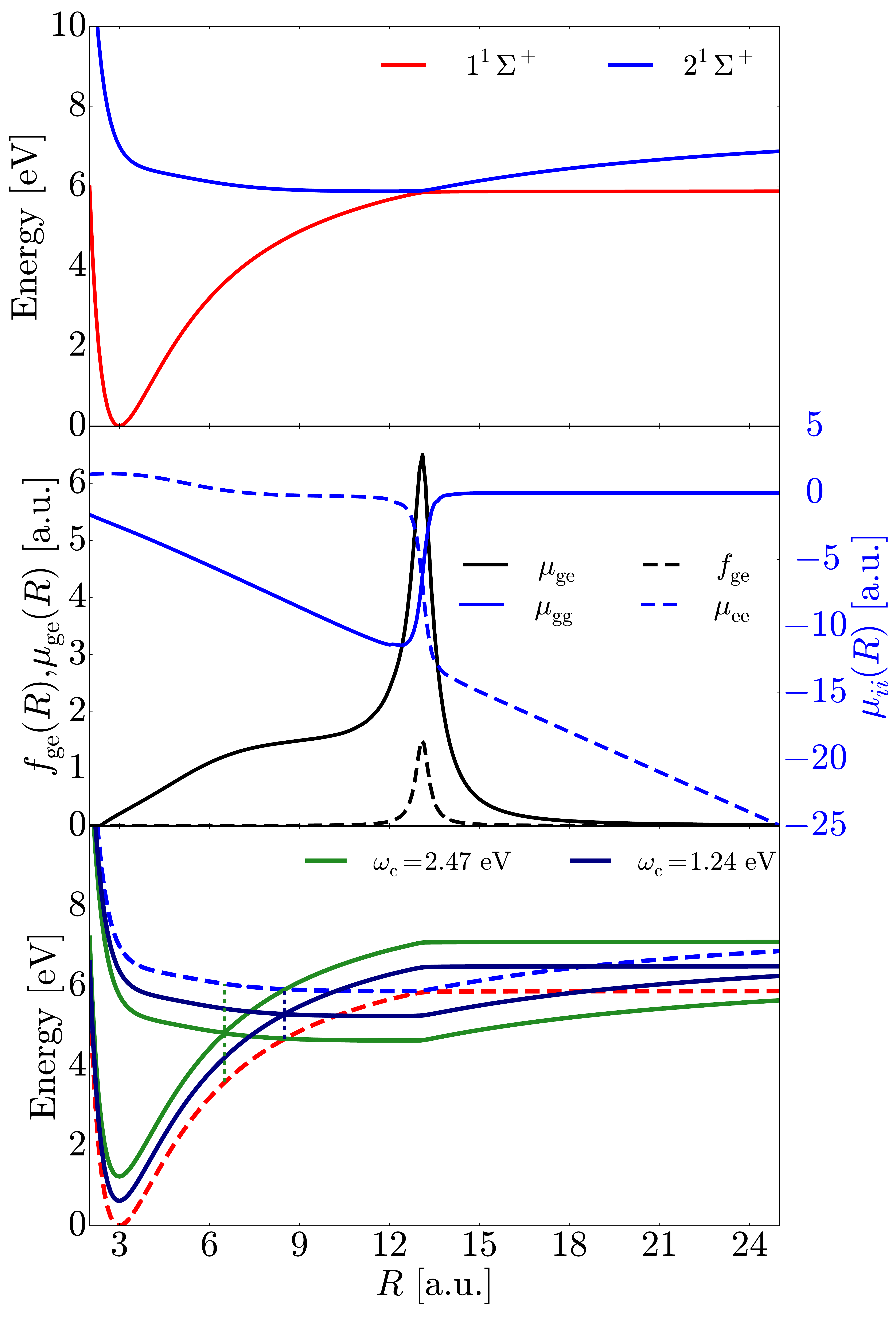}
\caption{ \label{fig:FigureI} 
(Top panel) Potential energy curves of the two lowest adiabatic states of LiF molecule
ground (g) $1\, ^1\Sigma$ state and excited (e) $2\; ^1\Sigma$ state. These states display
a (covalent-ionic) avoided crossing at $R=13.1$ a.u. The zero energy is set at the bottom of the ground state and the equilibrium internuclear distance is $R=3$ a.u.;
(Middle panel) Diagonal dipolar moments $\mu_{gg}(R)$ and $\mu_{ee}(R)$, and transition dipolar moment $\mu_{ge}(R)$, along with
the non-adiabatic coupling $f_{ge} (R)$. Note the different scales  in the $y$-axis;
(Lower panel) Modified potential energy curves for the light dressed states due to two different cavity mode frequencies, $\omega_c=1.24$ eV and
$\omega_c=2.47$ eV, that show LIC located at $R = 8.5$ a.u. and $R = 6.5$ a.u., respectively (internuclear distances at which the 
detuning $\Delta=[E_e(R)-E_g(R)]/\hbar]-\omega_c$ is zero). Adiabatic potential energy curves $E_{g}(R)$ and $E_e(R)$ are also introduced for comparison
(dashed lines).}
\end{figure}

In this fully ab initio study the PECs and couplings (dipolar plus non-adiabatic) must be computed at the highest level available
before any molecular dynamics calculation. For that purpose we use the electronic structure package MOLPRO \cite{MOLPRO_brief}.
For the PECs, we have performed a complete active space (CAS) calculation of the two lowest $^1\Sigma$ states of LiF using a
multiconfigurational self consistent field (MCSCF) method,  then followed by multi-reference configuration interaction (MRCI) method, 
using the {\em aug-cc-pVQZ} basis set, with the LiF molecule oriented along the $z$ axis. In addition,  the non-adiabatic couplings, 
and dipole moments (diagonal and non-diagonal) were computed with the wave functions at the MRCI level. In the case of the 
non-adiabatic couplings, these were calculated with MOLPRO using a finite differences method for the derivatives involving the 
MRCI wave function \cite{Werner1981}. Figure \ref{fig:FigureI} shows the adiabatic PECs of the two lowest $^1\Sigma$ states for LiF, 
the diagonal dipole moments $\mu_{gg}(R), \mu_{ee}(R)$ of these two states (LiF has a permanent dipole),
the non-diagonal transition dipole moment $\mu_{ge}(R)$ and the non-adiabatic couplings $f_{ge}(R)$. The latter indicates
a strong coupling between the two states at the avoided crossing located at $R=13.1$ a.u.  Due to this anticrossing the 
electronic wave functions exchange their character and, consequently, their adiabatic molecular properties (like $\mu_{gg}$
and $\mu_{ee}$) follow the behavior seen in figure \ref{fig:FigureI}. It is important to note that whereas $\mu_{ee}$ is close to zero
from the Franck-Condon region to the avoided crossing region, $\mu_{gg}$ is clearly dominant against $\mu_{ge}$, so that diagonal
couplings cannot be disregarded in the dynamics.

It is known that the presence of radiation (in the semiclassical or quantum form) induces light induced crossings (LIC) between
states coupled by the radiation \cite{}. In the quantized version, within the Jaynes-Cumming model, the new two dressed states (eigenstates of the
total Hamiltonian) show a splitting in the form \cite{Kowalewski2016}:
\begin{equation}
E_{\pm} = \frac{E_g(R) + E_e(R)}{2} \pm \frac{\hbar}{2} \Omega_{n_c} (R),
\end{equation}
where $E_g(R)$ and $E_{e}(R)$ are the adiabatic energies and $\Omega_{n_c}$ is the Rabi frequency, i.e.,
\begin{equation}
\Omega_{n_c}(R) = \sqrt{ 4 [\lambda (R)]^2 (n_c+1) + \Delta_c(R)}
\end{equation}
Here $\lambda$ is the interaction coupling factor,  $n_c$ stands for the Fock photon number in the cavity and $\Delta_c = [ E_e(R)-E_g(R)]/\hbar -\omega_c$
is known as detuning. In the case of LiF, the dressed curves are plotted in figure \ref{fig:FigureI} for two cavity mode frequencies, $\omega_c=1.24$ eV and $\omega_c=2.47$ eV.
The light induced crossing (LIC) appears at the internuclear distance $R_{\rm LIC}$ where the energy difference between the adiabatic curves, $[ E_e(R)-E_g(R)]/\hbar$,  
equals the cavity mode frequency $\omega_c$ (in figure \ref{fig:FigureI} we do not show the more irrelevant second crossing at a larger internuclear distance). This means 
that the chosen mode frequency $\omega_c$ turns out to be an interesting control parameter in the subsequent dynamics by selecting the position of the LIC.
It is worth noting that in our 
time dependent approach we expand the total wave packet in terms of the unperturbed Hamiltonian (non-interacting molecular adiabatic states plus field states),
so that the effect of the LIC that appears in the stationary dressed state picture must be represented by 
the adiabatic basis and the interaction within the MCTDH dynamics. In other words, while the dipolar and non-adiabatic couplings are directly introduced in the dynamics,
the LIC is not, but nevertheless it arises as a dynamical interaction. From figure $\ref{fig:FigureI}$ it is clear that the light induces 
effective dressed potentials in the $R-$direction, with shifted turning points and modified energetics. In our MCTDH method, we stress that the fully entangled wave packet 
dynamics takes place in potential energy surfaces given by 
$E_{g/e}(R)  + \frac{1}{2}\omega^2_c x^2  + \chi \omega_c \sqrt{2\hbar}  \mu_{gg/ee}(R)x$,
and the dynamics cannot be separated in the $x$ and $R$ coordinates.  

\begin{figure}[h!]
\centering
\includegraphics[width=0.5\textwidth]{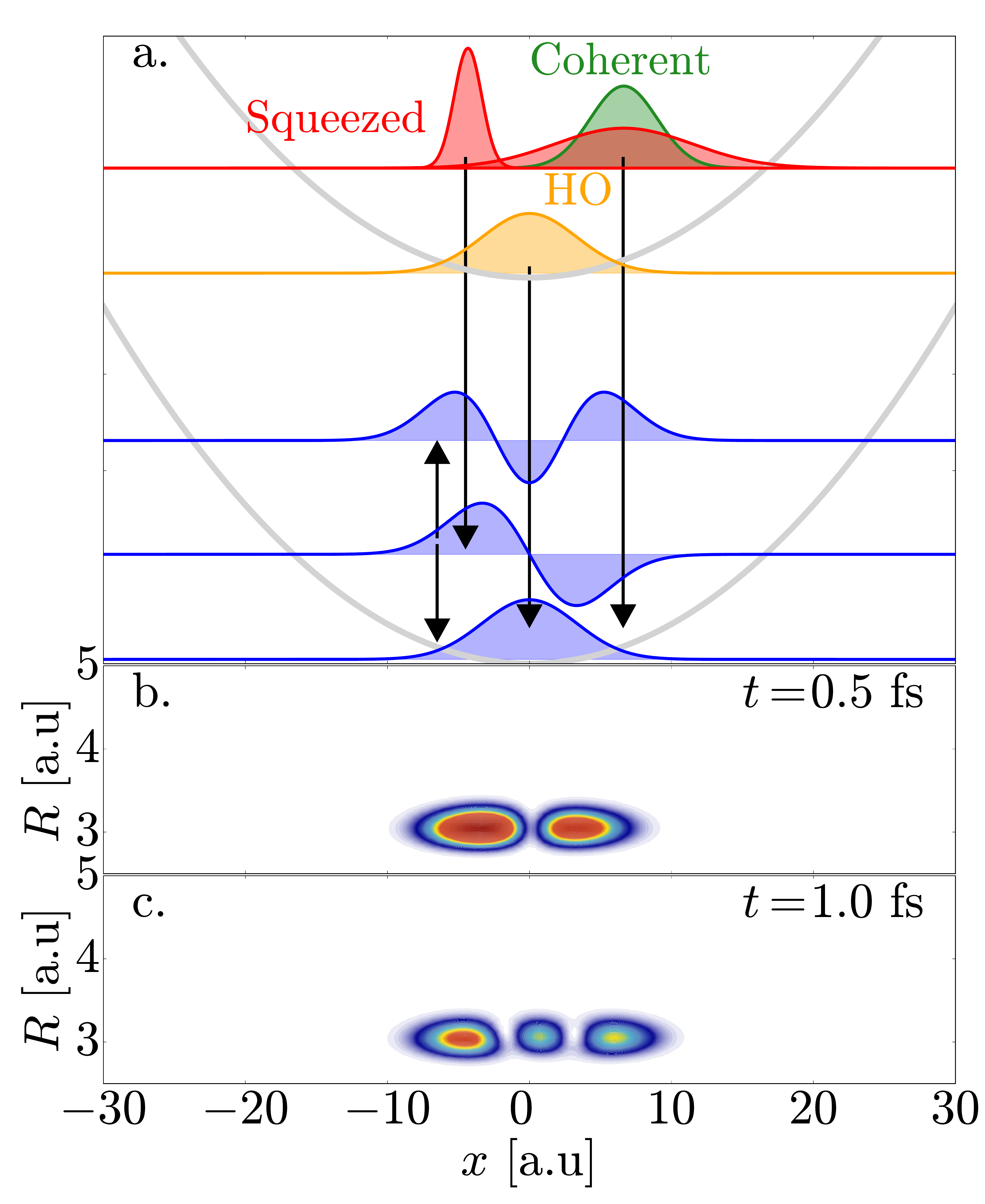}
\caption{ \label{fig:FigureII} 
(a) Scheme of different quantum states of radiation prepared on the upper harmonic oscillator (HO) potential with cavity mode frequency $\omega_c=2.47$ 
(associated to the $2 ^1\Sigma$ state),  namely, a Fock state or HO eigenstate $|n \rangle$ (shadow in blue or yellow), a coherent state $|\alpha \rangle$ (shadow in green) 
and a squeezed coherent state $|\alpha, \xi \rangle $ (in red) (the latter at two different times at which it narrows or widens). Vertical arrows
indicate transitions from the prepared radiation quantum states in the upper HO potential to the Fock states in the lower HO potential (also with the same 
frequency $\omega_c$ and associated to the $1 ^1\Sigma$ state). Also, since in the Frank-Condon region $\mu_{ee}(R) \sim 0$ intra-state dipole transitions 
are dominant only in the lower HO potential due to the large $\mu_{gg}$ value.
(b) Entangled photonic-nuclear probability density that appears (empty at $t=0$) in the lower PES at $t=0.5$ fs, along the vibrational mode
coordinate $R$ and the harmonic oscillator mode coordinate $x$, for an initial vacuum Fock state and $\omega_c=2.47$ eV. 
(c) Same as (b) but at a time $t=1$ fs. The $g$-wave packet thus evolves from a one-node to a two-node splitting in less than 1 fs (see text).
}
\end{figure}

The initial state must be prepared as an eigenfunction of the unperturbed Hamiltonian (molecule plus field), expressed 
as a direct product of the vibrational ground state of the $1\; ^1\Sigma$ state and the ground state corresponding to the harmonic oscillator of the cavity.
To initiate the molecular dynamics in the upper $2\; ^1\Sigma$
state we assume a prompt electronic excitation (produced by an intense laser field at $t < 0$) from the vibrational ground state in the $1\; ^1\Sigma$ state.
Due to the Frank-Condon approximation, it reduces to place the $v''=0$ vibrational state of the $1\; ^1\Sigma$ state in the $2\; ^1\Sigma$ PEC. Note that 
concerning NaI, the initial wave packet $v''=0$ was artificially shifted \cite{Kowalewski2017, Csehi2017a,Csehi2017b}
 to the right of the equilibrium distance to avoid a dominant
direct dissociation through the $2\; ^1\Sigma$ state. This is not required in our present study of LiF. The initial molecular wave packet  is obtained by using a 
direct diagonalization of the nuclear Schr\"odinger equation in a basis or using the relaxation method with imaginary time. 
Concerning the initial state of the HO, which must be also placed in the, so to speak, {\em upper} HO potential, 
our idea is to introduce different initial states of quantized light in the cavity, namely, i) a Fock state, ii) a coherent state an iii) a squeezed state, 
being the most common pure radiation states in quantum optics \cite{Gerry2005,Walls2009}. We do not enter here in the feasibility of the 
experimental preparation of these radiation states within the cavity. Our aim is to show that the molecular photodynamics can be quite different according to
the selected quantum state of radiation and to the choice of radiation mode frequency $\omega_c$. 
For the purpose of illustration, we have introduced in figure \ref{fig:FigureII} the three types of radiation quantum states.
The vacuum Fock state $| 0 \rangle$ corresponds to the ground state of the HO with potential $\frac{1}{2}\omega_c^2 x^2$ and it is an stationary state. 
A coherent state $| \alpha \rangle$ with a chosen average number of photons $\bar{n}$ is a superposition of Fock states with average energy $\bar{E}=\sum^{\infty}_{n=0}  P_n(\bar{n}) E_n(\omega_c)$,
where $P_n=e^{-\bar{n}} \frac{\bar{n}^n}{n!}$ and $E_n=\hbar\omega_c(n+\frac{1}{2})$. The coherent state is not an energy eigenstate and it bounces back and forth within the HO potential 
without losing its shape. The squeezed coherent state $ | \alpha, \xi \rangle$ is another superposition of  Fock states but, at variance, its wave packet, in addition to moving
back and forth in the HO potential, {\em breathes} (it widens and narrows by modifying its width and height). Due to the interaction term that couples the $e$ and $g$ states,
$\chi \omega_c \sqrt{2\hbar}  \mu_{ge}(R)  x$, the initial Fock states in the upper HO potential (linked to the $2\; ^1\Sigma$ molecular state) couple to Fock states of the 
lower HO potential (linked to the $1\;^1\Sigma$ state).

Although the cavity photodynamics of NaI has been recently treated in Ref. \cite{Kowalewski2016, Kowalewski2017, Csehi2017a, Csehi2017b}, we would like to provide a 
more comprehensive account of the inner workings of the underlying mechanisms that give shape to the time-dependent populations and the dissociation yields in the similar
molecule LiF.  We want to address the problem assuming that the cavity can be filled with different types of quantum states of radiation, namely, Fock states, coherent states and squeezed coherent states, which are known to have very different properties.

\subsection{Photodynamics with Fock states.}

In the case of Fock states, the initial total wave packet is a direct product $\chi^g_{v=0} (R) \times \psi_{n''=0} (x)$, where $\chi^g_{v=0} (R)$ is the vibrational ground state
of the $1\; ^1\Sigma$ state and $\psi_{n''=0}(x)$ is the ground state of the HO with mass unity and frequency $\omega_c$ (i.e., the radiation
Fock vacuum state $|0\rangle$ represented in coordinate space). As a rule we have used a radial box for the vibrational mode with $R \in [1.6,60]$ a.u. and another box for the
HO mode with $x \in [-x_{\rm max}:x_{\rm max}]$ a.u., with $x_{\rm max}=30-70$, large enough indeed to accommodate the three different quantum states of radiation under study, 
specially for the wide squeezed state. In our MCTDH calculation, we normally use $n_R=n_x=10-15$ single particle functions to represent both the molecular state in the nuclear box 
and the radiation state in the radiation box. The number of primitive basis is $N_R=1169$ for the molecular vibrational mode and $N_x=200-350$ for the radiation mode.

We study the evolution of this initial state subject to the molecule-cavity interaction, using different
cavity mode frequencies. These frequencies are selected for the matching $\omega_c=[E_e(R)-E_g(R)]/\hbar$  (detuning zero) in the LiF potential
energy curves (see figure \ref{fig:FigureI}) in order to choose the position $R_{\rm LIC}$ of the LIC, where the dressed curves are nearly degenerate
(see Theory section).  In table \ref{tab:TableI} we include the cavity mode frequencies (and the corresponding wavelengths) used in this work, that range from the visible
to the IR sectors in the electromagnetic spectrum. We also indicate in this table the corresponding position of the LIC in the PECs of LiF. Note that for 
$\omega_c \to 0$ the LIC approaches the position of  the non-adiabatic coupling (NAC), not induced by the radiation, located at $R_{\text{NAC}}=13.1$ a.u.  in LiF.

\begin{table}[t]
\begin{centering}
\begin{tabular}{l l l l l l l}
$\lambda_c$ [nm]       	& 400  		& 500  	& 600  	& 700  	& 1000 	& 2250 \tabularnewline \hline 
$\omega_c$ [eV]   		& 3.01 		& 2.47 	& 2.09 	& 1.76 	&  1.24  	& 0.54	\tabularnewline \hline 
$R_{\text{LIC}}$ [a.u.]     	& 5.8 		& 6.5		& 7.0 	& 7.5		&  8.5	&10.5	 \tabularnewline \hline 
\end{tabular}
\par\end{centering}
\caption{ Set of cavity mode wavelengths $\lambda_c$ and  frequencies $\omega_c$ used in this work. The cavity frequency 
determines the position of the light induced crossing (LIC) at the internuclear distance $R_{LIC}$, where the detuning $\Delta=[E_e(R)-E_g(R)]/\hbar-\omega_c$
is zero.
\label{tab:TableI}}
\end{table}

\begin{figure}[h!]
\centering
\includegraphics[scale=0.20]{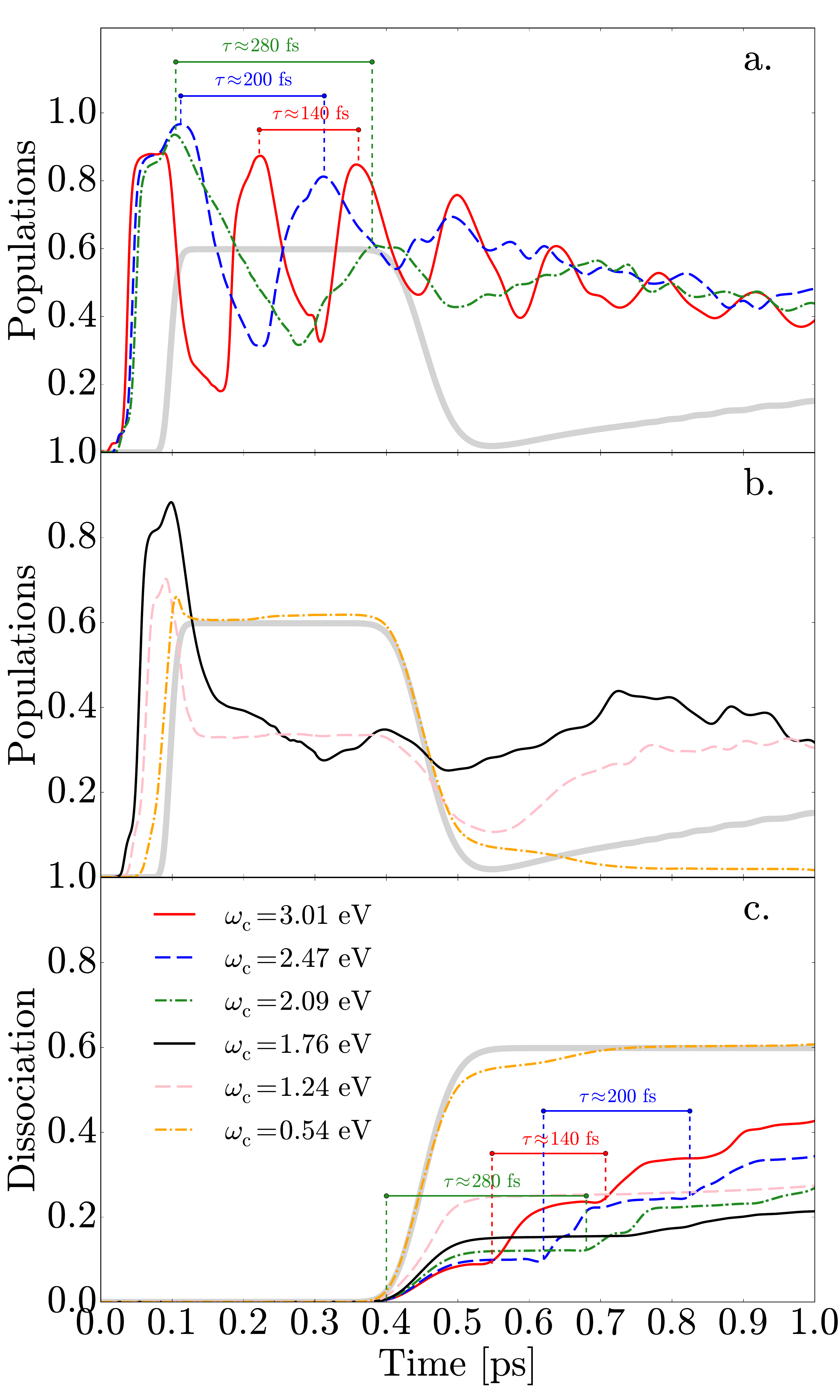}
\caption{ \label{fig:FigureIII} 
(a) Time-dependent population corresponding to the adiabatic ground state $1 ^1\Sigma$ for the molecule LiF initially fully excited in the $2 ^1\Sigma$ and 
coupled with the radiation present in a cavity in the form of the Fock vacuum state $|0\rangle$. Probabilities are included for three 
different cavity mode frequencies $\omega_c=3.01$ eV (red solid line), $\omega_c=2.47$ eV (blue dashed line), $\omega_c=2.09$ eV 
(green dashed-dotted line), with $\chi=0.05$.
The population for isolated LiF molecule (without a cavity) is also included as a reference (thick grey line) and its rapid decay after 400 fs is due to the 
presence of the absorbing potential situated at $R=58$ a.u. The periodicity between consecutive maxima is indicated in the figure.
(b) Same as (a) but for smaller cavity mode frequencies: $\omega_c=1.76$ eV (black solid line), $\omega_c=1.24$ eV (grey dashed line)  and
$\omega_c=0.54$ eV (orange dashed-dotted line). Again, the probability of cavity-undressed LiF is also included as a reference.
(c) Total dissociation probability (fully represented by the flux escaping asymptotically above the dissociation threshold of
 the ground state $1 ^1\Sigma$), for the same six cavity mode frequencies $\omega_c$ present in (a) and (b).}
\end{figure}

The time-dependent populations for the $g$-state (linked to $1\; ^1\Sigma$) are plotted in figure \ref{fig:FigureIII}(a)
and  \ref{fig:FigureIII}(b). The dissociation yield, that comes from a portion of the $e$-wave packet transferred to the $g$-state
through the different couplings (cavity and non-adiabatic) are included in figure  \ref{fig:FigureIII}(c).
For the largest $\omega_c$ values (3.01, 2.47 and 2.09 eV), the  $g-$population shows 
short-time oscillations with a clear periodicity (the larger the frequency, the shorter the period). These
oscillations tend to fade out at the long time limit. This effect is produced by the interaction with the 
cavity radiation, since in absence of the cavity the only effective coupling corresponds to the NAC 
located at $R=13.1$ a.u. and in this case the $1\; ^1\Sigma$ population 
does not show oscillations (see also figure \ref{fig:FigureIII}).  
In absence of radiation, the latter NAC is an efficient coupling 
(see figure \ref{fig:FigureI}),  since  more than half of the population of the $e$-wave packet is transferred to 
the $g-$state from the $e-$state at $t \sim 100$ fs. Eventually all this
population transferred to the ground state $1\; ^1\Sigma$ leads to dissociation (see figure \ref{fig:FigureI} and figure \ref{fig:FigureIII}(c)). 
All dissociation yields are computed with the flux crossing a complex absorbing potential located at $R=58$ a.u., at the edge of our spatial $R$-grid 
of size 60 a.u.  This absorbing layer is reached by the $g$-wave packet in less than 400 fs. Note that the $1\; ^1\Sigma$ population increases 
steadily again from $t=500$ fs. At this time the $e$-wave packet (reflected from the repulsive right side of the 
$2\; ^1\Sigma$ PES) meets again the NAC and a portion is transferred once more to the 
ground $1\; ^1\Sigma$ state (the increasing spreading of the $e$-wave packet causes the slow increase 
of the population during the subsequent visits to the NAC).  Since the $e$-wave packet bounces back and forth 
in the $2\; ^1\Sigma$ potential and it crosses the NAC region twice within a vibrational cycle (two transfers 
to the ground state),  at $t \to \infty$ the whole population initially located in the upper $2 ^1\Sigma$ state will eventually lead 
to full dissociation of the LiF molecule through the $1 ^1\Sigma$ state. In our MCTDH simulations (due to computational limitations), with a final time of 1 ps for the propagation, the 
$e$-wave packet does not complete a full vibrational cycle in the $2\; ^1\Sigma$ state, so that the highest dissociation yield at this time is 0.6. 

Back to the field dressed case, the oscillations in the populations tend to disappear as $\omega_c$ decreases 
(see cases with $\omega_c = 1.76,1.24$ and $0.54$ eV in figure \ref{fig:FigureIII}). As $\omega_c \to 0$, the position of the LIC approaches 
the location of the NAC and the effect of the cavity radiation vanishes. Also, note that the periodicity $\tau$ of the oscillations present in the 
time-dependent populations  ($\tau=140$ fs for $\omega_c = 3.01$ eV, $\tau=200$ fs for $\omega_c=2.47$ eV, and $\tau=280$ fs for $\omega_c=2.09$ eV
in figure \ref{fig:FigureIII}(a)) also emerges in the dissociation yields. Whereas for the undressed LiF molecule the only 
presence of a NAC produces a single burst of dissociation within the time window of the first ps, the effect of the cavity field is to 
produce a train of dissociating wave packets with a $\omega_c$-dependent periodicity $\tau$. For the three larger values of 
$\omega_c$, after each period $\tau$, the dissociation yield increases by  a quantity which is almost independent of the cavity mode 
frequency $\omega_c$ (i.e., same amounts but at different time delays). This is because the dissociative $g$-wave packet always comes from the
$e$-wave packet through the NAC with the same transfer rate, at variance with the variable transfer rate at the LIC, since this interaction strength 
itself depends upon $\omega_c$. For the three smaller values of $\omega_c$ (1.76, 1.24 and 0.54 eV) the behavior of the dissociation 
yield represents a tug of war between the LIC and the NAC effects. The closer proximity of the LIC to the NAC makes the NAC physics dominant,
but still there is an interfering  LIC contribution, which is less effective for smaller $\omega_c$ values.

\begin{figure*}[ht!]
\centering
\includegraphics[width=0.8\textwidth]{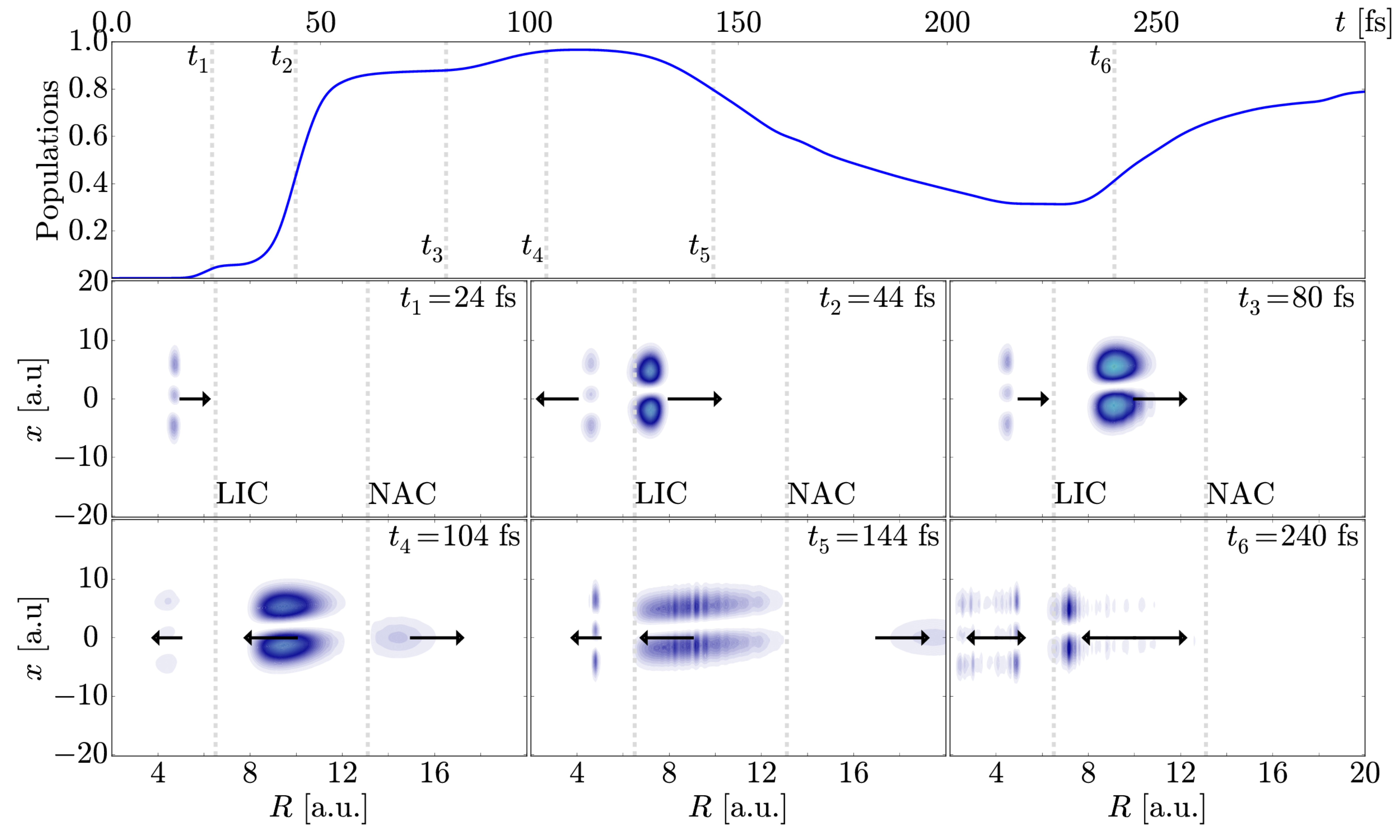}
\caption{ \label{fig:FigureIV} Snapshots of the time-dependent evolution for the probability density of the entangled radiation-molecule wave packet 
moving in the potential energy surface $E_g(R) + \frac{1}{2}\omega_c x^2 + \chi \omega_c \sqrt{2\hbar} \mu_{gg}(R) x$ (see Eq. \ref{eq:coupledTDSE})
of LiF. The plot corresponds to a cavity mode frequency $\omega_c=2.47$ eV, interaction factor $\chi=0.05$ and the radiation is set up initially as the vacuum 
Fock state $|0\rangle$. The times $t_i$ for the snapshots are connected with different features present in the population of the $g-1 ^1 \Sigma$ state (upper panel). 
Arrows indicate the direction of motion of the wave packet and its size is related to the magnitude of its momentum. Vertical lines within the snapshots indicate
the internuclear distance $R=6.5$ a.u. for the location of light induced crossing (LIC) and the internuclear distance $R=13.1$ a.u. for the non-adiabatic crossing (NAC).  }
\end{figure*}

\begin{figure*}[ht!]
\centering
\includegraphics[width=0.8\textwidth]{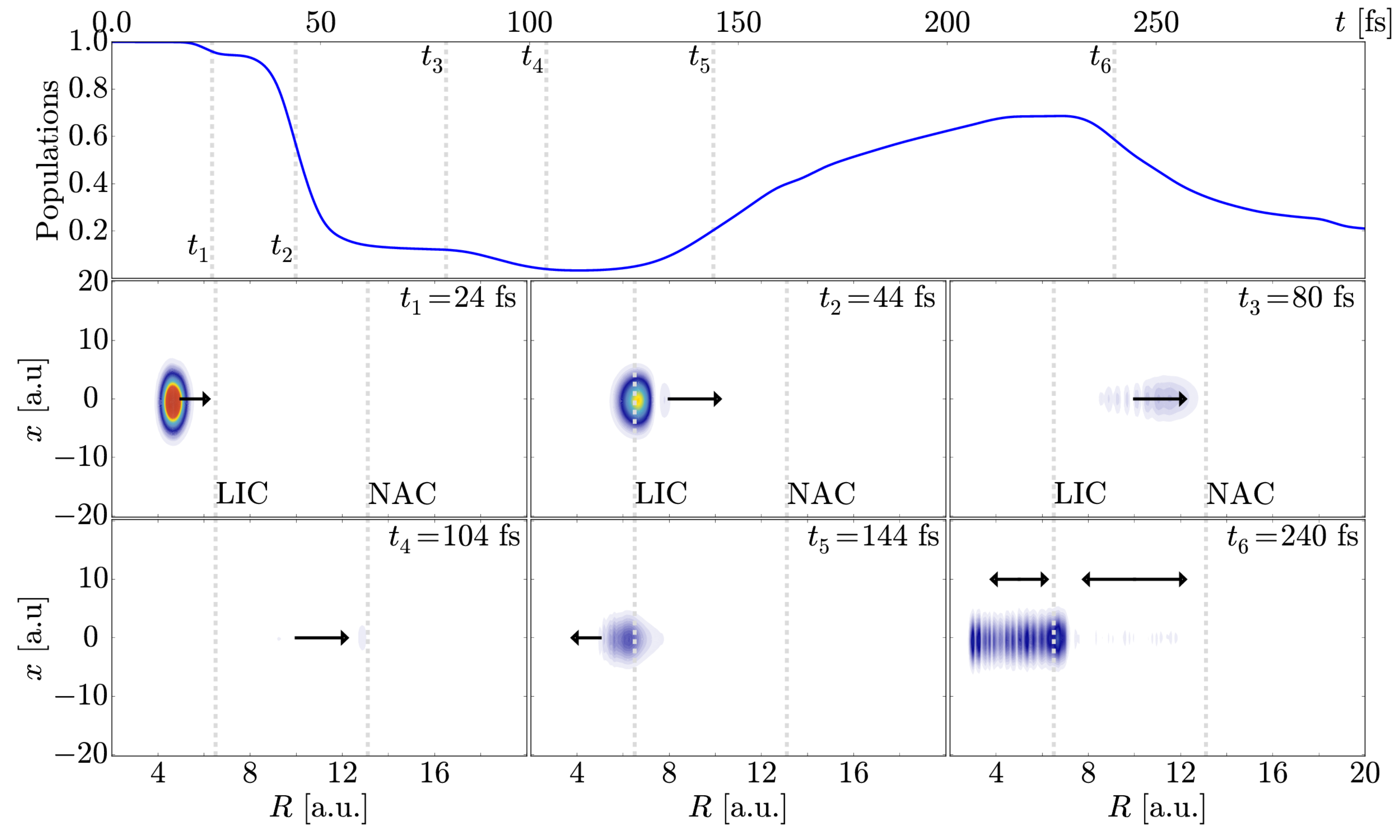}
\caption{ \label{fig:FigureV} Same as figure \ref{fig:FigureIV} but for the population and time-dependent 
probability density of the $e-2 ^1 \Sigma$ state (initially populated at $t=0$) moving within the potential energy surface given by 
$E_e(R) + \frac{1}{2}\omega_c x^2 + \chi \omega_c \sqrt{2\hbar} \mu_{ee}(R) x$. Snapshots are taken at the same times as in figure \ref{fig:FigureIV} for comparison. }
\end{figure*}

To understand the inner workings of the cavity photodynamics of LiF, we choose to analyze in detail the prototypical case with 
cavity frequency $\omega_c=2.47$ eV.  At this initial total energy the dissociation process comes through 
the population transfer from the upper $e$-PES to the lower $g$-PES. Then, the  time dependent populations 
in these two states as well as the evolution of the probability densities for the corresponding $g$- and $e$-wave packets are plotted
in figure \ref{fig:FigureIV} and figure \ref{fig:FigureV}, respectively.

The initial unentangled photonic-nuclear $e$-wave packet  at $t=0$ in $2\; ^1\Sigma$ is approximately a direct product of 
two Gaussians of different widths. Its compact  form is quite robust during dissociation (see figure \ref{fig:FigureV}). However, the underlying 
dynamics is better understood following the $g$-wave packet density in figure \ref{fig:FigureIV}. Firstly, at t=24 fs we appreciate that the $g$-wave 
packet splits in  three portions along the $x$ (radiation mode) coordinate. In fact, this splitting happens to occur below 1 fs.  In figure $\ref{fig:FigureII}$ 
we have included the $g$-wave packet density  at $t=0.5$ and $t=1$ fs, which shows a distribution with one node and two nodes, respectively. 
The interaction term $\chi \omega_c \sqrt{2\hbar}  \mu_{ij}(R)  x$ causes transitions, so to speak,  between vibronic states
in both the molecular (through $\mu_{eg}(R)$) and the HO subsystems (through $x$). According to 
figure \ref{fig:FigureI}, the dipole moment $\mu_{eg} (R)$ is not zero in LiF for $R > R_e$ ($R_e \sim 3$ a.u. is the equilibrium 
distance), so that transitions between the upper ($e$) and lower ($g$) HO states take place promptly. According to HO 
dipole selection rules its ground state $\psi^e_{n'=0}$ makes a sudden and dominant transition to  $\psi^g_{n''=1}$ in the lower 
HO potential (hence the $g$-wave packet is created with a single node at the beginning). Moreover, since whereas the diagonal
dipole moment $\mu_{ee}$ is almost negligible in $R_e < R < 12$ a.u., the diagonal dipole moment $\mu_{gg}$ is not (see figure \ref{fig:FigureI}). 
Indeed, its magnitude is larger than the transition dipole moment $\mu_{eg}$ in the same region. This means that the interaction term 
$\chi \omega_c \sqrt{2\hbar}  \mu_{gg}(R)  x$ also produces intra-state HO transitions within the $g-$state, from the previously populated $n''=1$ to $n''=0$ and
mainly $n''=2$ (see figure \ref{fig:FigureII}), which generates a two-node $g$-wave packet extended in the $x$ coordinate whose shape remains robust
during its propagation in the confined zone $0 < R < R_{LIC}$, the inner region of the dressed potential $E_{-} (R)$ 
(figure \ref{fig:FigureI}).  These $e-g$ transitions at short $R<R_{\rm LIC}$ are always present because the cavity interaction 
is always switched on but it is feeble because is quite out of resonance (the detuning is zero only at the position of the LIC). 

The cavity interaction depletes the upper state population when the $e$-wave packet  reaches the LIC,  located at $R=6.5$ a.u. for $\omega_c=2.47$ eV. 
At this internuclear distance the transition dipole moment $\mu_{eg}$ is much more effective and the $e$-wave packet passing through the LIC  produces 
a one-node $g$-wave packet in the lower PES which propagates with an inherited momentum. While this newly LIC-created wave function 
approaches the NAC region, the two-node $g$-wave packet has almost completed a vibrational cycle (approximately $\tau=56$ fs) 
from $t=24$ to $t=80$ fs, inside the dressed $1\; ^1\Sigma$ state. Similarly, the one-node $g$-wave packet bounces back and forth confined in the region 
$R_{\rm LIC} < R < R_{\rm NAC}$ with a period $\tau \sim 200$ fs.

The second enhancement of the ground state population is due to the NAC at $R=13.1$ a.u.. Between 
$t=80$ fs and $t=104$ fs the remaining $e$-wave packet almost disappears because its density is fully transferred to the 
$g-$state at $t=104$ fs (see time $t_4$ in figure \ref{fig:FigureIV} and figure \ref{fig:FigureV}). This latter portion of $g$-wave packet finally leads to dissociation. The 
middle one-node $g$-wave packet, after reflection from the $1\; ^1\Sigma$ turning point near the NAC, approaches the LIC 
at $t \sim 144$ fs. Again, the efficient LIC drives a large portion ($\sim 0.6$) of the one-node $g$-wave packet back to the upper 
$e$-PES, where a  nodeless and  extended $e$-wave packet emerges with its inherited momentum. In other words, from 
$t=$144 to $t=$240 fs the initial $e$-wave packet is partially reconstructed thanks to the LIC. Therefore, from this time onward, 
the full mechanism is again repeated, but with a smaller initial population. These cyclic mechanism explain the main features present
in the populations in figure \ref{fig:FigureIII}(a). 

In conclusion, there are notorious features introduced by the cavity interaction, namely,
i) multiple splittings of the wave packet along the internuclear distance $R$ (at the LIC and at the NAC) 
and the  presence of nodes along the $x$ coordinate due to radiation mode transitions.
ii) portions of the the density remain bound and confined in the regions
$0 < R < R_{\rm LIC}$ and $R_{\rm LIC} < R < R_{\rm NAC}$
iii) dissociation fluxes can be produced with a higher repetition rate and controllable delay (by choosing the cavity mode frequency) 
than with undressed molecules.

Note that the wave packet dynamics in figure \ref{fig:FigureIV} does not preserve a parity symmetry $x \to -x$ for the densities.
This is because the PES for the ground state reads $E_g(R) + \frac{1}{2}\omega_c x^2 + \chi \omega_c \sqrt{2\hbar} \mu_{gg}(R) x$ 
and the non-zero negative value of $\mu_{gg}$ for $R$ produces an additional linear potential in the $x$ direction (see art graphic
in the Abstract with the 3D plot of the $g-$ and $e-$PES).
As mentioned above we have included the adiabatic PECs and interaction couplings (cavity and non-adiabatic terms) so that
the effective dressed states and LIC between them must be reproduced by our basis dynamically. From the time-dependent
densities we learn of the virtual presence of the LIC; our strong cavity transitions occur at the $R_{\rm LIC}$ positions predicted by the
dressed state picture (see table \ref{tab:TableI}). Moreover, the two-node $g$-wavepacket confined at short $0 < R < R_{LIC}$ 
shows different reflections and vibrational periods for each chosen cavity mode frequency $\omega_c$. For instance, 
for $\omega_c=3.1$ eV this portion of the $g-$wave packet is reflected at $R \sim 4$ a.u.; for $\omega_c=2.47$ the turning point is 
at $R \sim 5$ a.u. (see figure \ref{fig:FigureIV};  and for $\omega_c=2.09, 1.76$ and $1.24$ eV this inner $g$-wave packet travels up 
to $R \sim 6$, 7 and 8 a.u., respectively, and then is reflected back. This is an indication of the existence of a movable LIC between 
virtual dressed PES.

\subsection{Photodynamics with coherent states}

\begin{figure*}[ht!]
\centering
\includegraphics[scale=0.20]{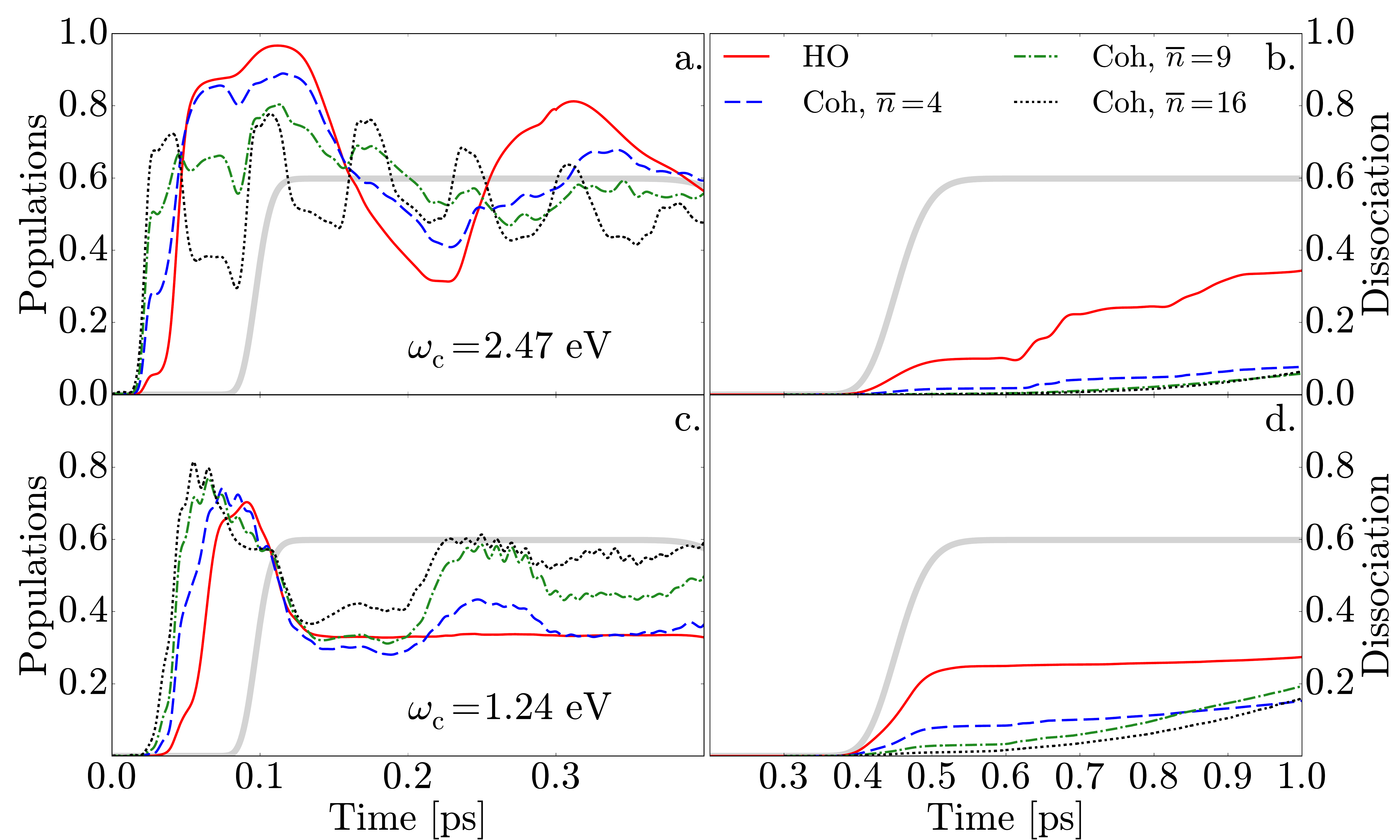}
\caption{ \label{fig:FigureVI} 
(Left panels) Time-dependent population corresponding to the adiabatic ground state $1 ^1\Sigma$ for the molecule LiF initially fully excited in the $2 ^1\Sigma$ and 
coupled with the radiation present in a cavity in the form of a coherent state $|\alpha \rangle$, with different average number of photons $\bar{n}=|\alpha|^2=4$ (blue dashed line), 
9 (green dotted-dashed line) and 16 (black dotted line). Probabilities are included for two different cavity mode frequencies $\omega_c=2.47$ eV (a) and 
$\omega_c=1.24$ eV (c), with $\chi=0.05$. The probabilities for the undressed dynamics (grey thick line) and for the photodynamics with a Fock vacuum state
(red solid line) are also included as a reference.
(Right Panels) Total dissociation probability (represented by the flux escaping asymptotically above the dissociation threshold of
 the ground state $1 ^1\Sigma$), for the  cavity mode frequencies $\omega_c=2.17$ eV (b) and $\omega_c=1.24$ eV (d), 
 associated to the probabilities in (a) and (c), respectively.}
\end{figure*}

Now we assume that the optical cavity is filled with a coherent radiation.
The coherent state of light in the position space representation is given by \cite{Gerry2005}
\begin{align}
\psi_{_\alpha}(x)=\left(\frac{1}{2\pi [\Delta x_{_\alpha}]^2}\right)^{1/4}
e^{-\frac{1}{4} \left( \frac{x-\langle x \rangle_{_\alpha}}{\Delta x_{_\alpha}} \right)^{2}}
 e^{i\langle p \rangle_{_\alpha}x} 
\end{align}
where $\Delta x_{_\alpha}=\sqrt{\frac{\hbar}{2\omega_c}}$ corresponds to the width, and
the wave packet is centered initially at $\langle x \rangle_{_\alpha}=\sqrt{\frac{\hbar}{2\omega_c}}(\alpha+\alpha^{\ast})$ in the $x$-spatial grid
and with an initial  momentum given by $\langle p \rangle_{_\alpha}=-i\sqrt{\frac{\hbar\omega_c}{2}}(\alpha-\alpha^{\ast})$.
$\alpha=|\alpha| e^{i\psi}$ is a complex number that characterizes the coherent state in the complex phase space and it is related to
the average number of photons in the cavity through $\bar{n}= |\alpha|^2$. Here we have chosen $\psi=\pi/4$
and also selected different values for the average number of photons, namely, $\bar{n} = 4, 9$ and $16$. Now the initial total wave packet 
is the direct product $\chi^g_{v=0} (R) \times \psi_{\alpha} (x)$.
Then the choice for $\bar{n}$ and the cavity mode frequency $\omega_c$ determine the initial location and 
momentum of the coherent state, which is readily represented in our radiation grid with $x \in [-50,50]$ a.u.
The larger the value for $\bar{n}$, the larger the average energy, $\bar{E}=\sum_{n=0} P_n(\bar{n}) E_n(\omega_c)$, and the longer the oscillation
amplitude in the HO potential.  

\begin{figure*}[ht!]
\centering
\includegraphics[width=0.8\textwidth]{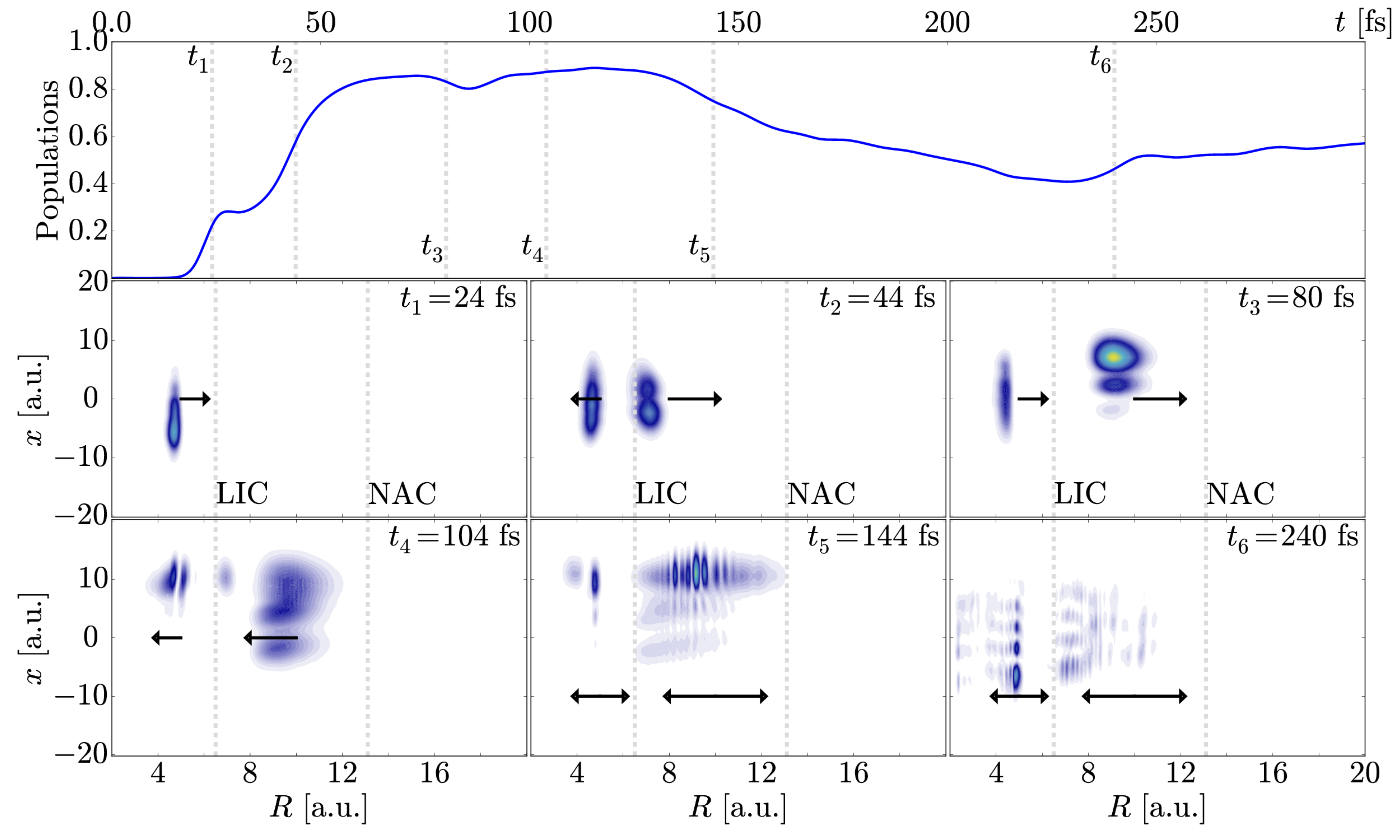}
\caption{ \label{fig:FigureVII} 
Snapshots of the time-dependent evolution for the probability density of the entangled radiation-molecule wave packet 
moving in the potential energy surface $E_g(R) + \frac{1}{2}\omega_c x^2 + \chi \omega_c \sqrt{2\hbar} \mu_{gg}(R) x$ (see Eq. \ref{eq:coupledTDSE})
of LiF. The plot corresponds to a cavity mode frequency $\omega_c=2.47$ eV, interaction factor 
$\chi=0.05$ and the radiation is set up initially as a coherent state $|\alpha\rangle$ with an average number of photons $\bar{n}=4$
and angular phase $\psi=\pi/4$. The times $t_i$ for the snapshots are connected with different features present in the population of the $g-1 ^1\Sigma$ state (upper panel). 
Arrows indicate the direction of motion of the wave packet and its size is related to the magnitude of its momentum. Vertical lines within the snapshots indicate
the internuclear distance $R=6.5$ a.u. for the location of light induced crossing (LIC) and the internuclear distance $R=13.1$ a.u. for the non-adiabatic crossing (NAC). 
 }
\end{figure*}

\begin{figure*}[ht!]
\centering
\includegraphics[width=0.8\textwidth]{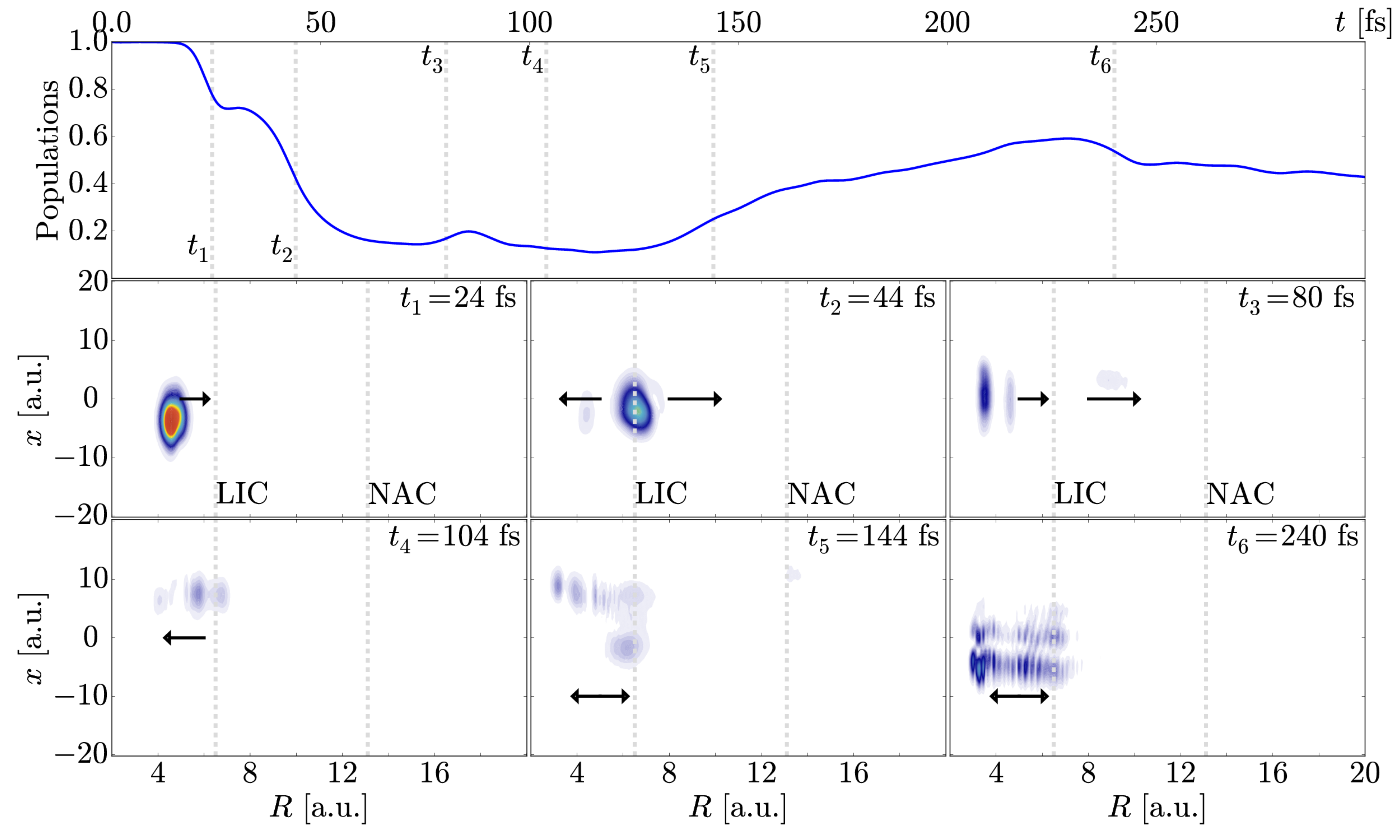}
\caption{ \label{fig:FigureVIII} 
Same as figure \ref{fig:FigureVII} but for the population and time-dependent 
probability density of the $e-2 ^1 \Sigma$ state (initially populated at $t=0$ with the radiation state as a coherent state) moving within the potential energy surface given by 
$E_e(R) + \frac{1}{2}\omega_c x^2 + \chi \omega_c \sqrt{2\hbar} \mu_{ee}(R) x$. Snapshots are taken at the same times as in figure \ref{fig:FigureVII} for comparison.
 }
\end{figure*}

In figure \ref{fig:FigureVI} we include the time-dependent $g-$state populations and the dissociation yields with radiation represented as a coherent state, and we choose
two cavity mode frequencies $\omega_c=1.24$ and 2.47 eV, both with three different average number of photons ($\bar{n}=4, 9$ and 16). For $\bar{n}=4$, according 
to $P_n$, the main Fock components in the coherent state extend from $n=0$ to $n=9$ with a maximum in $n=4$.
This means that the coherent state initially moves within the HO potential  with an average total energy higher to that of the vacuum Fock state. 
For $\omega_c=2.47$ eV and $\bar{n}=4$ the behavior of the $g$-population with $\bar{n}=4$ is similar to that of the Fock vacuum state. 
Actually, the coherent state is a displaced vacuum state, $|\alpha \rangle =D(\alpha) |0 \rangle$, with the same width. Nonetheless, we find some differences. At the short 
time range $20-50$ fs the $g$-population develops a
hump not present in the HO Fock case (see figure \ref{fig:FigureVI}) that increases further for $\bar{n}=9$ and 16. This $g$-population enhancement is caused by the larger
effective dipole transition in the HO mode along $x$. The coherent state is not an stationary state and it oscillates (preserving its shape) with a large amplitude between 
the turning points of the upper HO potential at the average energy $\bar{E} (\bar{n})$. This oscillation has a period $T=2\pi/\omega_c$ and its amplitude in larger for 
increasing $\bar{n}$. These long excursions of the coherent state to large values of $x$ makes that the effective dipole matrix elements between the 
moving coherent state and the HO Fock states of the $g$-state be more effective than between Fock states. This is reflected in the higher 
population transfer, even in the off-resonant region $0 <R < R_{\rm LIC}$ and also at $R_{\rm LIC}$. This higher population transfer must be  appreciated in the 
$g$-wave packet density (more intense than in the Fock state case); for instance, in the case $\omega_c=2.47$ eV and $\bar{n}=4$, see the first snapshot at time 
$t = 24$ fs included in figure \ref{fig:FigureVII}. This piece of $g$-wave packet has no node structures, given that the coherent state is a superposition of Fock states 
and this tends to wash out any sharp node corresponding to particular HO transitions. 

It must be noted that densities for $\omega_c=$2.47 eV in figures \ref{fig:FigureVII} 
and \ref{fig:FigureVIII} are frames at a fixed time, but these densities move upward-downward in the $x$ direction, with a period $T=1.67$ fs. 
For $\bar{n}=4$ the coherent state is initially centered at $x \sim 6$ a.u. and with a positive momentum $\langle p \rangle_{\alpha}$, then it reaches the right turning point where the dipole is 
maximum. There are no less than 20 oscillation cycles of the coherent state to reinforce the dipole transitions before the $e-$wave packet reaches the LIC. Also, at $t=44$ fs (see Figure \ref
{fig:FigureVIII}) a weak signal appears traveling in opposite direction to the $e$-wave packet moving across the LIC. It comes through the off-resonant cavity dipole transition, 
from  the population of the $g$-state already trapped in $0 <R < R_{LIC}$ ($t=44$ fs in Figure \ref{fig:FigureVIII}). This piece of $e-$wave packet moves in the 
upper $e$-PES with a different momentum than the $g-$wave packet moving in $0 < R < R_{\rm LIC}$ at $t=44-80$ fs. At $t=80$ fs the traveling piece 
of $g$-wave packet within the interval $0 < R < R_{\rm LIC}$ partially transfers to the $e-$state, thus producing two fringes of delayed $e-$wave packets with
slightly different momentum that eventually will interfere.  The LIC is so effective with coherent states  that the $e$-wave packet is almost emptied in 
the region $R < R < R_{\rm NAC}$ and little density survives to arrive at the NAC (see $t=80$ fs in figure \ref{fig:FigureVIII}).  This is why the dissociation 
yield in Fig. \ref{fig:FigureVI}(b) for any coherent state is much reduced with respect to the Fock case. Also, a large portion of the $g$-density confined in
$R_{\rm LIC} < R < R_{\rm NAC}$ is transfer back to the $e-$state between 144 and 240 fs.  Finally at $t=240$ fs the $e$-wave packet, partially
reconstructed with large interference structures, restarts the whole mechanism.

For $\bar{n}= 16$ the dynamical oscillatory pattern for the populations in figure \ref{fig:FigureVI} changes abruptly with a series a periodic structures separated by $\tau \sim 70$ fs.
In this case (not shown in the figures), it takes $\sim 34$ fs for the $e$-wave packet to reach the LIC, time at which the $g-$state population shows its first maximum peak. However, 
in this occasion  the $e-$wave packet do not cross the LIC, but it is reflected back for $R < R_{\rm LIC}$. Due to the large enhancement of the dipole moment for the coherent states
and even more for large $\bar{n}$, the $g-$population in exchanged with the $e$-state until the latter reaches the inner turning point where the $\mu_{eg} \sim 0$ 
(Figure \ref{fig:FigureI}). When the upper $e-$wave packet arrives at the equilibrium distance $R_e$, a new cycle repeats. 
Consequently, the sequence of peaks in figure \ref{fig:FigureVI} indicates the periodical
arrival of the $e$-wave packet to the LIC from $R_e$, and the full dynamics is confined in $0 < R < R_{\rm LIC}$, with the NAC playing no role. 
This mechanism is confirmed by changing the cavity mode frequency to $\omega_c=1.24$ (see figure \ref{fig:FigureVI}(c), a case for which the LIC is
located at $R_{\rm LIC}$ = 8.5 a.u.)  In this case the period for the confined motion in $0 < R <R_{\rm LIC}$ is longer, $\tau = 180$ fs, with the maximum humps indicating that
the $e-$wave packet approaches the LIC, makes the transition, and it is reflected back.

\section{Photodynamics with a squeezed-coherent state}

\begin{figure*}[ht!]
\centering
\includegraphics[width=0.75\textwidth]{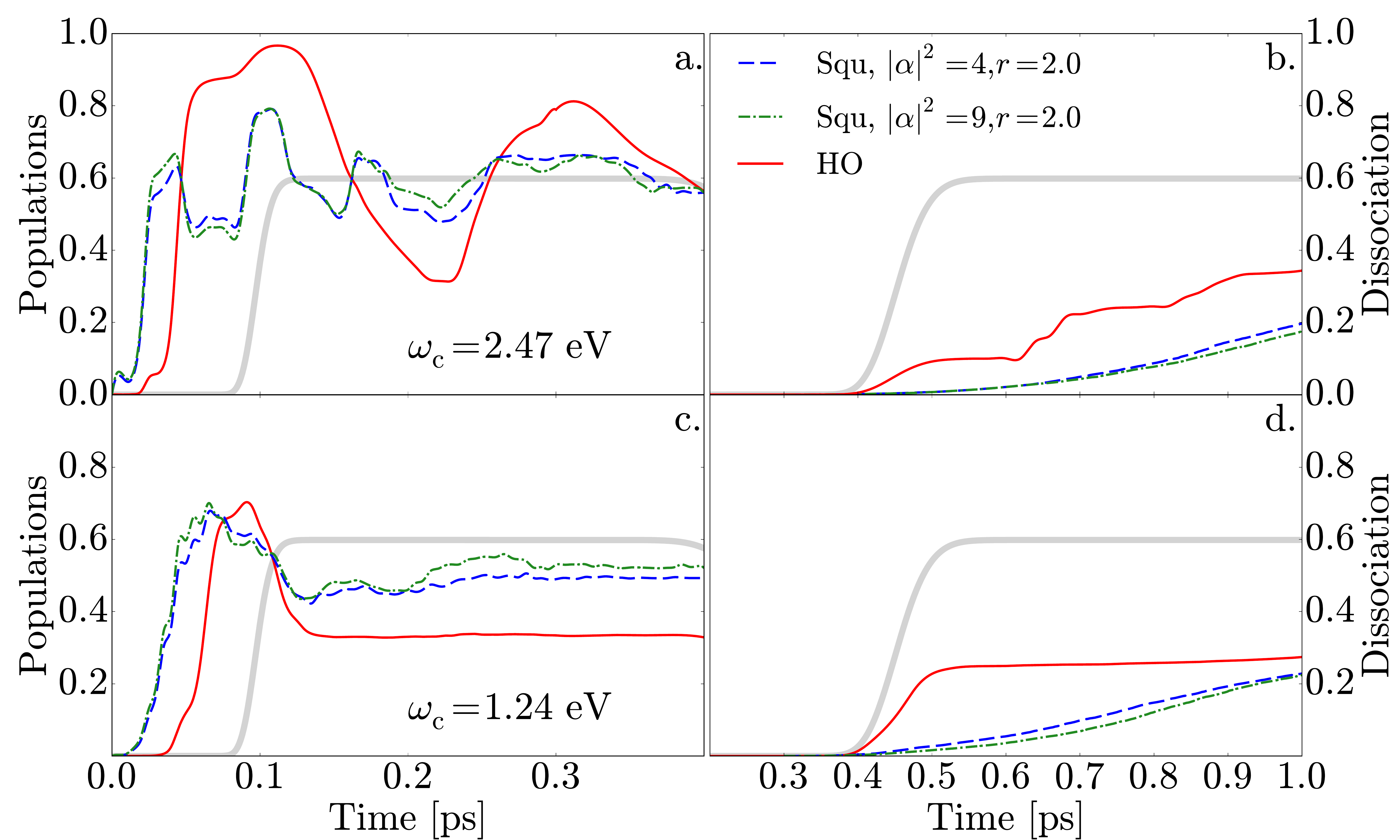}
\caption{ \label{fig:FigureIX}
(Left panels) Time-dependent population corresponding to the adiabatic ground state $g-1 ^1\Sigma$ for the molecule LiF initially fully excited in the $e-2  ^1\Sigma$ and 
coupled with the radiation present in a cavity in the form of an squeezed coherent state $|\alpha, \xi \rangle$, with two 
different values of $|\alpha|^2$=4 (blue dashed line) and $|\alpha|^2=9$ (green dotted-dashed line), 
squeezing parameter $r=2$ and angular phase angle $\psi-\theta/2=2\pi/5$,  are included for two different cavity mode frequencies $\omega_c=2.47$ eV (a) and $\omega_c=1.24$ eV (c). 
The probabilities for the undressed dynamics (grey thick line) and for the photodynamics with an initial Fock vacuum state 
(red solid line) are also included as a reference.
(Right Panels) Total dissociation probability (represented by the flux escaping asymptotically above the dissociation threshold of
 the ground state $1 ^1\Sigma$), for the  cavity mode frequencies $\omega_c=2.47$ eV (b) and $\omega_c=1.24$ eV (d), 
 associated to the probabilities in (a) and (c), respectively.
 }
\end{figure*}

The squeezed coherent state in the position representation is given by \cite{Moller1996}
\begin{widetext}
\begin{align}\label{eq:squeezed}
\psi_{\alpha, \xi}(x)=&\left( \frac{\omega_c}{\pi\hbar} \right)^{1/4}(\cosh r + e^{i\theta}\sinh r)^{-1/2} \times\exp\left[ -\left(\frac{x-\langle x \rangle_{_\alpha}}{2\Delta x_{\alpha}}\right)^{2} + \frac{i}{\hbar}\langle p \rangle_{_\alpha}\left(x-\frac{\langle x \rangle_{_\alpha}}{2}\right) \right],
\end{align}
\end{widetext}
with the width given by $\Delta x_{_\alpha}=\sqrt{\frac{\hbar}{2\omega_c}}\left( \frac{\cosh r + e^{i\theta}\sinh r}{\cosh r - e^{i\theta}\sinh r} \right)^{1/2}$
and the initial position $\langle x \rangle_{_\alpha}=\sqrt{\frac{\hbar}{2\omega_c}}(\alpha+\alpha^{\ast})$ and 
momentum $\langle p \rangle_{_\alpha}=-i\sqrt{\frac{\hbar\omega_c}{2}}(\alpha-\alpha^{\ast})$, with $\alpha=|\alpha| e^{i \psi}$. 
The squeezed state is characterized by the complex variable $\xi = r e^{i \theta}$, where $r$ is the squeezing parameter and $e^{i \theta}$ the squeezing phase.
We have fixed the angular difference $\psi-\theta/2=2\pi/5$. The coherent states have the property that the variances of the position $\hat{X}_1=(\hat{a} +\hat{a}^\dag)/2$ 
and momentum $\hat{X}_2=(\hat{a} -\hat{a}^\dag)/2i$ 
quadratures are equal, i.e., $\langle (\Delta \hat{X}_1)^2 \rangle= \langle (\Delta \hat{X}_1)^2 \rangle=1/4$ so that they minimize the position-momentum uncertainty product.
At variance, with a squeezed coherent state, one of the two variances can be chosen to be below 1/4 at the cost of increasing the other. A consequence of this property
is that the spatial wave packet associated to the squeezed coherent state breaths (it narrows when $\Delta X_1 < 1/4$ but it widens when $\Delta X_1 > 1/4$). This effect
is appreciated if the state in Eq.~\ref{eq:squeezed} is propagated in time (see figure \ref{fig:FigureII} for an illustrative picture). At the time of maximum expansion, the spatial allocation of the squeezed 
state may require the use of large grids $x \in [-70:70]$, specially for large $\bar{n}$. The average number of photons for a  squeezed coherent state is given by
$\bar{n}=|\alpha|^2 + \sinh^2 r$. In our computations we have chosen $r=2$ as squeezing parameter and $|\alpha|^2=4$ and 9. In these cases, the squeezing
character dominates over the coherent one in the contribution to $\bar{n}$.

\begin{figure*}[ht!]
\centering
\includegraphics[width=0.80\textwidth]{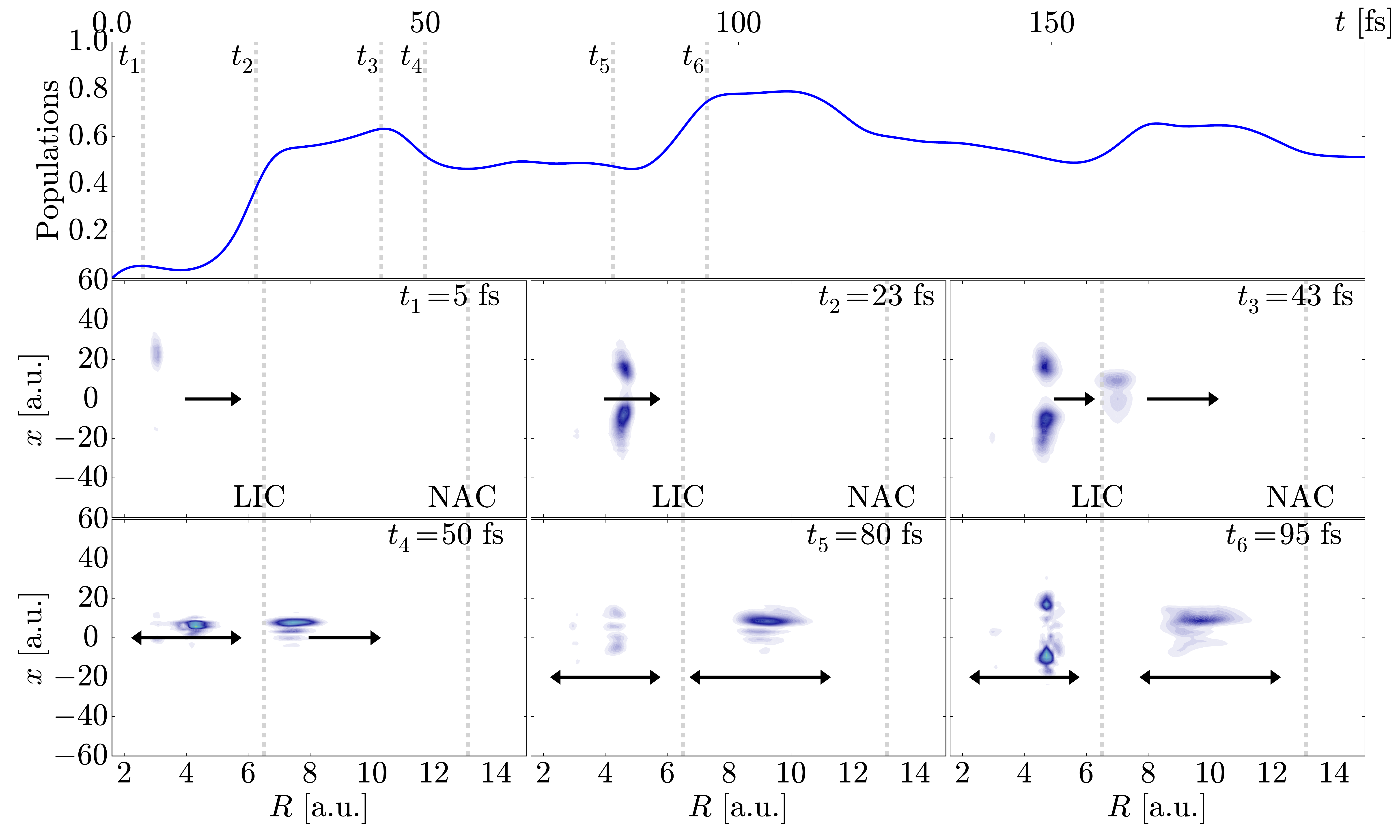}
\caption{ \label{fig:FigureX}  
Snapshots of the time-dependent evolution for the probability density of the entangled radiation-molecule wave packet 
moving in the potential energy surface $E_g(R) + \frac{1}{2}\omega_c x^2 + \chi \omega_c \sqrt{2\hbar} \mu_{gg}(R) x$ (see Eq. \ref{eq:coupledTDSE})
of LiF. The plot corresponds to a cavity mode frequency $\omega_c=2.47$ eV, interaction factor 
$\chi=0.05$ and the radiation is set up initially as a squeezed coherent state $|\alpha ,\xi \rangle$ (with an average number of photons $\bar{n}=4$
and coherent and squeezing phase $\psi=\theta=0$, and squeezing parameter $r=2$. The times $t_i$ for the snapshots are connected with different 
features present in the population of the $1 ^1\Sigma$ state (upper panel). 
Arrows indicate the direction of motion of the wave packet and its size is related to the magnitude of its momentum. Vertical lines within the snapshots indicate
the internuclear distance $R=6.5$ a.u. for the location of light induced crossing (LIC) and the internuclear distance $R=13.1$ a.u. for the non-adiabatic crossing (NAC).
}
\end{figure*}

\begin{figure*}[ht!]
\centering
\includegraphics[width=0.80\textwidth]{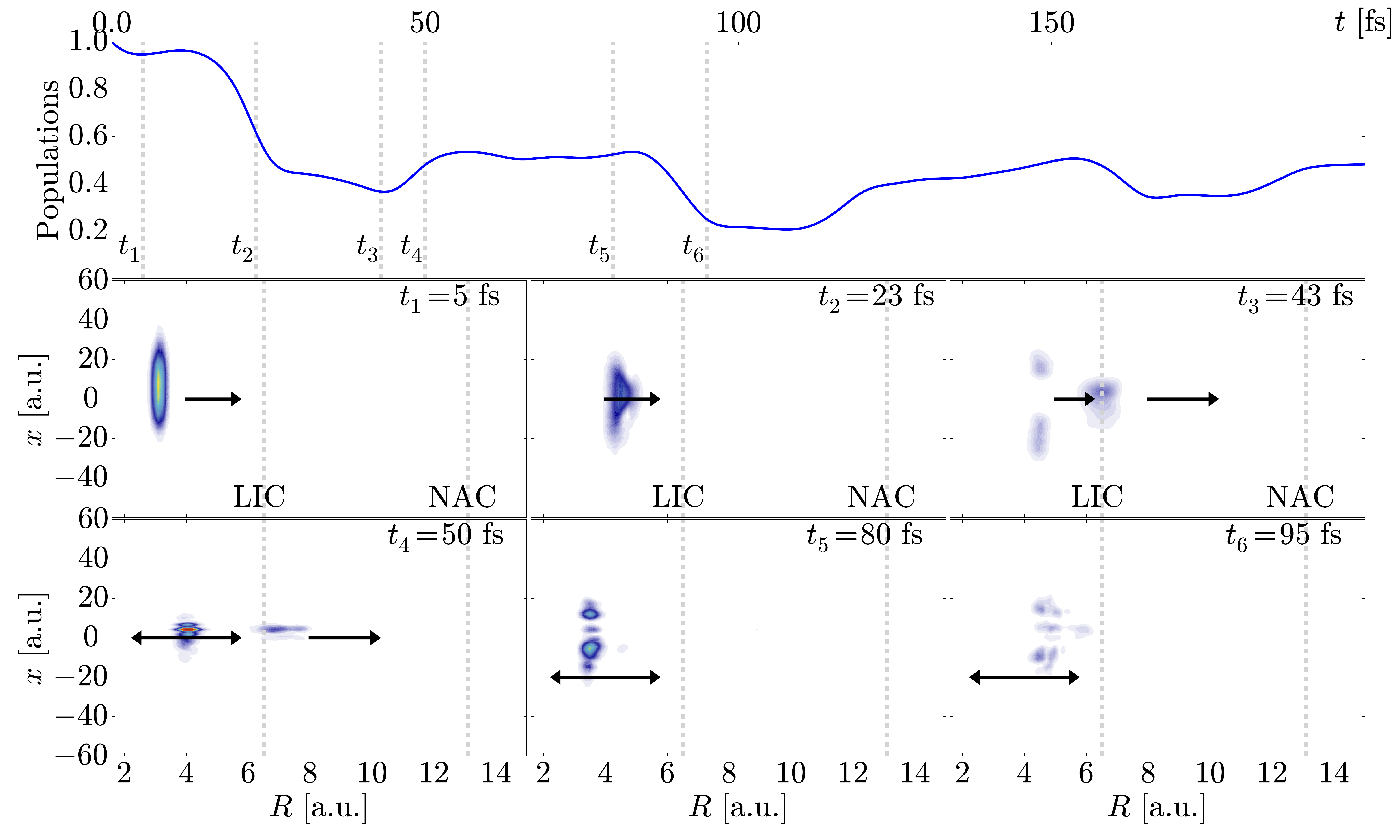}
\caption{ \label{fig:FigureXI}  
Same as figure \ref{fig:FigureX} but for the population and time-dependent 
probability density of the $2 ^1 \Sigma$ state (initially populated at $t=0$ with the radiation state as a squeeze coherent state) 
moving within the potential energy surface given by  $E_e(R) + \frac{1}{2}\omega_c x^2 + \chi \omega_c \sqrt{2\hbar} \mu_{ee}(R) x$. 
Snapshots are taken at the same times as in figure \ref{fig:FigureX} for comparison.}
\end{figure*}

In figure \ref{fig:FigureIX} we include the time evolution of the $g-$populations and dissociation probabilities 
for two cavity mode frequencies, $\omega_c=2.47$ and 1.24 eV. The patter of oscillations in the $g$-populations for both $|\alpha|^2=4$ and $|\alpha|^2=9$
are similar between them but also comparable to the $g$-populations for a coherent state with the largest value $\bar{n}=|\alpha|^2=16$. We learnt from the coherent radiation case 
that the increase of the dipole moment as $\bar{n}$ increases was the responsible for these rapid oscillations with a short period of 70 fs. The same situation appears 
with squeezed coherent states but with smaller values of $|\alpha|^2$. The cause is that the transition dipole moments for the squeezed state are larger than for the coherent
state. The squeezed coherent state for $\omega_{c}=2.47$ eV, $|\alpha|^2=4$ and $r=2$ breaths twice the coherent oscillatory period $T=1.67$ fs. If one calculates the
time dependent dipole moment $D(t)=\int dx\; \psi^g_{n''=1}(x)  \;x  \;\psi_{\alpha, \xi} (x,t)$ between the Fock state $|1\rangle$ and the evolving squeezed coherent 
state $|\alpha, \xi \rangle$ one finds a maximum magnitude of 2.45 a.u., while the same calculation for a moving coherent state provides 1.70 a.u.
Consequently, it means that the squeezing introduces large effective dipole couplings even with smaller values of $|\alpha|$. Thus the cavity couples
the two states mainly in the internuclear region $0 < R < R_{\rm LIC}$ and the LIC is again more crucial than the NAC in the dynamics. The
off-resonant cavity interaction in $R_e < R < R_{\rm LIC}$ and the LIC itself depopulates the $e-$state (see figures \ref{fig:FigureX} and \ref{fig:FigureXI} for 
$t=5-43$ fs). The $e$-wave packet  is almost no longer present in the region $R_{\rm LIC} < R < R_{\rm NAC}$, and thus there is little dissociation. 
It transfers its population at the LIC to the $g$-state at $t \sim 43$ fs and the latter remains mostly confined in that region. On the other side, the density 
in the region $0 < R < R_{\rm LIC}$ is also confined but it is continuously exchanged between the $g-$ and the $e-$state, which results in oscillating populations, 
with a period corresponding to one cycle of vibration of the $g-$wave packet in this inner zone. It must be also noted that the oscillatory (with breathing) motion
along the cavity mode coordinate $x$ provokes a larger spreading and self-interferences in the wave packets. This spreading along $x$ 
(similar to the coherent state), which is not present in the Fock case, makes that $g/e$-wave packets visit a more extensive landscape of the PES. 
As mentioned above, both PES are not symmetric with respect to parity ($x \to -x$) due to the presence of diagonal dipole moments.  Its effect
is appreciated in the Fock case ($t=144$ fs in figure \ref{fig:FigureIV} where the $g-$wave packet in $R_{\rm LIC} < R < R_{\rm NAC}$ is biased toward positive $x$)
in the coherent case ($t=144$ fs in figure \ref{fig:FigureVII} ) and in the squeezed state (see $t=80$ fs in figure \ref{fig:FigureX}).

\section{Conclusions and perspectives}

Strong coupling in the light-matter interaction can be produced by confining a molecule within an optical cavity, so that the molecule 
interacts with the cavity quantum modes of light. 
In this work we have studied the photodynamics of LiF immersed within an optical cavity. 
We have conceived three scenarios for the quantum radiation states: Fock states, coherent states and squeezed coherent states as different 
expressions of quantum light. In other very recent studies it has been pointed out that there are no real radiation quantum effects when 
using Fock states  \cite{Csehi2017a,Csehi2017b} and that any special effect in the population dynamics or dissociation yields can be fully 
reproduced by using a semiclassical approach for the radiation-matter interaction, only just by fitting the correct intensity of the classical pulse 
to the constants of the interaction terms in the cavity-matter interaction. We believe that this is not a general case and that the conditions under 
which a quantum field could be replaced by his counterpart classical field are still an open problem. With this work we hope to contribute to the 
understanding of the dynamics of a simple polar diatomic molecule when subject to cavity quantum radiation. We have shown that the response of the system can be
drastically different when using different forms of quantum states for the cavity. For that purpose we have performed fully ab initio calculations for both
the structure and photodynamics of the molecule, using state of the art tools to calculate the temporal evolution of the
 entangled (matter plus radiation) wave packet. Strictly speaking, the molecular populations should be calculated from the reduced
 density operator  $\hat{\rho}^{mol}$ for the molecule (computed by tracing over the total density operator corresponding to 
 the entangled total wave packet). In this respect along this work we have always shown populations for the light-dressed system
 (light plus molecule), but we judged them sufficient to illustrate our purposes.

In LiF photodynamics three regions along the internuclear distance are clearly identified: i) the inner region $0 < R < R_{\rm LIC}$ 
, ii) the middle region $R_{\rm LIC} < R < R_{\rm NAC}$ and iii) the outer region $R > R_{\rm NAC}$. We have compared the different dynamics of the three 
quantum states by fixing the same interaction strength $\chi$ for all of them. When the radiation state is prepared as a Fock vacuum
state, both LIC and NAC play fundamental roles. In the Fock case we have analyzed vibronic transitions among Fock states  
between different electronic states or within the same (ground) electronic state. The latter effect arises only if diagonal dipole
moments are introduced in the dynamics.  The onset of the LIC is generated (in our formalism) by the dynamical interaction
and its position $R_{\rm LIC}$ in the MCTDH simulations coincides with that expected from dressed state simple calculations.
The portion of $e$-wave packet transferred at $R_{\rm LIC}$ to the $g-$state in the middle region $R_{\rm LIC} < R < R_{\rm NAC}$ remains
confined but partially helps to reconstruct the initial $e-$wave packet because of successive transfers across the LIC in the reverse 
direction.

The confinement of the density within the inner region  $0 < R < R_{\rm LIC}$ is a paramount effect of the interaction with the cavity. 
The density of this  confinement is larger for coherent and squeezed states of light due to the enhancement of the effective dipole moment 
along the $x$ direction together with a non-negligible $\mu_{eg}(R)$ molecular dipole moment  in the inner region. 
In coherent and squeezed states 
the effects of cavity radiative couplings in $0 < R < R_{LIC}$ dominate over the effect produced by the NAC.
This strongly modify the LiF photodynamics and cavity Rabi oscillations in the inner region dominate the whole dynamics,
with dissociation mostly suppressed.  When the cavity is in a vacuum Fock state, 
we expect LiF to deliver  a train of dissociative matter wave packets separated by a controllable delay between them.  
As a future perspective, the cavity mode frequency $\omega_c$ and the type of quantum state of radiation prepared in the cavity 
can be set up as usable factors that can modify and control the dynamics of simple polar diatomic molecules, which in the 
undressed case is only dominated by a single NAC.

\begin{acknowledgements}

We thank the GFIF-GFAM-UdeA computer cluster for generous allocation of computer time. 
This work was supported by  financial support from Vicerrector\'{\i}a de Investigaci\'on 
(Estrategia de Sostenibilidad) at Universidad de Antioquia and from Departamento Administrativo
de Ciencia, Tecnolog\'{\i}a e Innovaci\'on (COLCIENCIAS, Colombia), under Grant No. 111565842968. 
D.P. is thankful to the CaPPA project (Chemical and Physical Properties of the Atmosphere),
funded by the French National Research Agency (ANR) through the PIA (Programme d'Investissement d'Avenir) 
under Contract No. ANR-10-LABX-005 (Labex CaPPA) for financial support.
JFTG thanks Colciencias for a PhD grant Cr\'edito Condonable. JLSV thanks 
the scientific {\em grandparentship} of Manuel Y\'a\~nez and Otilia M\'o along many years
of dedicated support.

\end{acknowledgements}





\bibliography{biblio}

\end{document}